\documentclass[floatfix,prl,aps,twocolumn,superscriptaddress,10pt]{revtex4-2}

\usepackage{silence}
\usepackage[utf8]{inputenc}
\usepackage[T1]{fontenc}
\usepackage[english]{babel}
\AtBeginDocument{\renewcommand{\selectlanguage}[1]{}}
\usepackage{color}
\usepackage{xcolor}
\usepackage{amsmath}
\usepackage{amsfonts}
\usepackage{hyperref}
\usepackage{amsthm}
\usepackage{bm}
\usepackage{float}
\usepackage{textcomp}
\usepackage{upgreek}
\usepackage{textgreek}
\usepackage{setspace}
\usepackage{amssymb,graphicx}
\usepackage[percent]{overpic}
\hypersetup{
  colorlinks = true,
  allcolors= blue,
}
\usepackage{tikz,pgfplots}
\usepackage{blkarray}
\usepackage{booktabs}
\usepackage{array}
\usepackage{multirow}
\usepackage{caption}
\DeclareCaptionJustification{fulljustified}{\leftskip=0pt \rightskip=0pt \parfillskip=0pt plus 1fil}
\usepackage{braket}

\usetikzlibrary{patterns}
\usetikzlibrary{fadings}
\usetikzlibrary{decorations.pathmorphing,patterns}
\tikzset{snake it/.style={decorate, decoration=snake}}
\usetikzlibrary{arrows, decorations.markings}
\pgfplotsset{compat=1.14}
\tikzset{
vecArrow/.style={
  thick,
  decoration={markings,mark=at position
   1 with {\arrow[scale=2,thin]{open triangle 60}}},
  double distance=1.4pt, shorten >= 10.5pt,
  preaction = {decorate},
  postaction = {draw,line width=1.4pt, white,shorten >= 4.5pt}
  },
innerWhite/.style={
  semithick,
  white,
  line width=1.4pt,
  shorten >= 4.5pt
  }
}
\DeclareMathAlphabet{\mathsfit}{\encodingdefault}{\sfdefault}{m}{sl}
\SetMathAlphabet{\mathsfit}{bold}{\encodingdefault}{\sfdefault}{bx}{sl}

\begin{document}

\title{
    Magnons in solids with correlated atomic and nanoscale disorder
    }

\author{Alex Moorhouse}
\affiliation{Department of Applied Physics and Applied Mathematics, Columbia University, New York, US}
\author{Joseph Barker}
\affiliation{School of Physics and Astronomy, University of Leeds, Leeds, UK}
\author{Michele Simoncelli}
\email{michele.simoncelli@columbia.edu}
\affiliation{Department of Applied Physics and Applied Mathematics, Columbia University, New York, US}

\begin{abstract}
Atomic defects and disorder affect both the microscopic magnon excitations and the macroscopic physical properties of real-world magnetic materials. 
The rationalization of this relation has hitherto remained elusive, because of the intertwined challenges of simulating with quantum accuracy atomic- and nanoscale inhomogeneities and correlations, and their 
effect on the magnon spectral function determining observables such as spin stiffness.
Here, we introduce a theoretical and computational framework to address these challenges. 
First, we develop the Coupled Optimization of SpIN Environment (COSINE) algorithm for efficiently relaxing the
atomic structure and magnetic configuration of a solid with defects or disorder to a local energy minimum. Then, we discuss how to efficiently obtain exchange interactions accounting for local compositional, structural, and magnetic-moment inhomogeneities. Finally, we develop a paraunitary unfolding approach for Bogoliubov bosons, which allows us to predict the magnon spectral function in materials with any magnetic order, including systems with canting and full, partial, or no magnetic-moment compensation. 
We showcase our approach in a representative defective ferrimagnet for generality, and then combine it with state-of-the-art spin-informed machine-learning potentials to quantitatively predict how nonuniformities in the spatial distribution of vacancy defects control blurring of effective magnon bandstructures.
\end{abstract}

\maketitle
\section*{INTRODUCTION}

Defects, dopants, and structural distortions affect the physical properties of many magnetic materials \cite{silveyra, gutfleisch_21_century}. 
These deviations from perfect crystalline long-range ordering in magnetic materials are subtle, because they may exist on different length scales \cite{herzer_modern_soft, dagotto_colossal, hono_strategy}, and in different degrees of freedom, such as the atomic positions or magnetic moments \cite{simonov_designing_disorder, binder_spin_glasses}.
From a technological viewpoint, it is very useful to understand and predict how these deviations from perfect long-range order---often broadly termed ``disorder''---affect macroscopic properties such as magnetization, spin stiffness, and Curie temperature.

Specifically, disorder in a magnet may be perturbative, altering the system only slightly compared to the pristine case, or sufficiently strong to induce qualitative changes in how it responds to temperature and external magnetic fields (e.g., transition from ferromagnetic to paramagnetic behavior \cite{PhysRevLett.91.137202, Jeon_2027}). Additionally, disorder may affect multiple degrees of freedom simultaneously or only a subset; for example, a magnet can feature long-range order in the magnetic moments while being disordered in composition and atomic positions \cite{jiang_feni_sro, zhang_hea_afm}, or possess inhomogeneous nanostructure \cite{herzer_modern_soft}. 
Meanwhile, crystallographically ordered solids may lack long-range magnetic order
due to frustrated interactions \cite{ramirez_spin_ice, castelnovo_spin_ice} or thermal fluctuations \cite{als_nielsen_euo_eus}. Superparamagnetic materials can lack both magnetic ordering and long-range structural order beyond the length scale of their nanograins \cite{bedanta_supermagnetism}, and lattice distortions can coexist with magnetic frustration in spin glasses \cite{booth_y2mo2o7}.
Overall, the magnetic properties of solids are influenced in a non-trivial way by 
the extent of disorder in structural, compositional, and magnetic moment degrees of freedom \cite{gautam, li_wu_soft_magnets}, and a long-standing challenge is to understand and predict how.

Early experimental studies in the 1960s tackled this challenge using neutron scattering, and found that increasing concentrations of nonmagnetic dopant atoms in Fe--Si alloys decrease the spin stiffness \cite{shirane_1965, shirane_1968}, which is minimized in an alloy of two magnetic metals in equal stoichiometric ratio \cite{menzinger_1968}. 
A nearly contemporaneous theoretical study employed a spin-wave model accounting for doping inhomogeneities with an expansion around the mean crystalline lattice \cite{murray}, assuming an uncorrelated distribution (i.e., not capable of describing clustered inhomogeneities at the nanoscale). This theoretical study showed a quantitative discrepancy compared to subsequent, more precise measurements in Fe--Si alloys \cite{antonini}.
More recent spin-lattice dynamics simulations employed the pairwise exchange approximation---exchange parameterized as a strictly pairwise function, dependent only on the distance between two atoms and their chemistry, so it is not sensitive to the local atomic environment (LAE) surrounding the two atoms---to further investigate how uniform (uncorrelated) random distributions of vacancies affect magnons in iron, finding that increasing the vacancy concentration leads to a decrease in magnon frequency and increase in magnon linewidth \cite{mudrick}, and found reasonable agreement with the pristine experimental stiffness in Ref.~\cite{loong}.

Pioneering efforts to go beyond uncorrelated descriptions focused on describing nanoscale inhomogeneities \cite{chakraborty,buczekSpinWavesDisordered2018a}, 
and to keep this tractable they relied on state-of-the-art approximations for the treatment of atomic-scale inhomogeneities.
Overcoming these approximations requires accounting for lattice distortions, local magnetic-moment variations, and anisotropic or LAE-dependent exchange interactions, and then their interplay with nanoscale inhomogeneities. Hitherto, this has been prohibitively challenging for quantitative simulations because of two reasons: there is extensive evidence \cite{chapman_nonheisenberg, PhysRevB.38.11718, chapman_mlip, kormann_2008, hasegawa_1983, izyumov_1966, sabiryanov, hammerschmidt, kaushikFiniteTemperatureEnhancementMagnetic2026, rubanTemperatureInducedLongitudinalSpin2007, shallcrossInitioEffectiveHamiltonian2005} that they are intertwined in a non-trivial way, and they are beyond the reach of first-principles spin-polarized Density-Functional Theory (DFT) \cite{marzari_electronic_structure, horton_magnetic_order, li_deeph_magnetic} which effectively limits simulations to hundreds-of-atoms ($\lesssim$ 1 nm) cells. In such cells, periodic boundary conditions hinder the realistic simulation of spatial correlations.

In summary, capturing the complex interplay between inhomogeneities in atomic structure, magnetic moments, and exchange parameters, as well as the role of correlated nanoscale disorder has so far remained prohibitively complex. Addressing such a problem is of fundamental importance, because errors arising from approximations at the atomic scale could fortuitously compensate for those at the nanoscale, suggesting apparent agreement with experiments while masking potentially powerful degrees of freedom for engineering and controlling the macroscopic properties of magnets \cite{Kaczmarek_2024}.
The control and patterning of defects or dopants in magnetic materials could provide a means of tuning the properties of magnons in doped magnets analogous to routes already established for phonons and thermal transport in composition-engineered solids \cite{thebaud,fiorentino,pazhedath_lapo4,garg_sige}. Practically, correlated structural vacancies may be introduced by irradiation \cite{hepburn, beeler}, mechanical stress \cite{li_cuo_vacancy_migration, gao}, or manufacturing processes \cite{liang_wustite}, and since the spin stiffness influences the Curie temperature \cite{pajda} and the width of magnetic domain walls \cite{marrowsSpinMixingSpinCurrent2004}, for example, quantifying these relationships and shedding light on related engineering opportunities is of significant practical importance.

In this work, we introduce a theoretical and computational framework to simulate magnons in non-homogeneous or defective materials explicitly using exchange interactions derived from spin-informed machine-learned interatomic potentials (SI-MLIPs), which have near first-principles quantitative accuracy. We address simultaneously the challenges of capturing the interdependence between lattice distortions, non-collinearities and longitudinal variations in the magnetic moments, and the sensitivity of exchange interactions to the LAE beyond pairwise interactions.
To achieve this goal, we introduce the theory and paraunitary unfolding approach to compute the spectral function of a system of bosons from their most general Bogoliubov Hamiltonian, and develop the computational infrastructure to evaluate this with quantitative accuracy using state-of-the-art SI-MLIPs \cite{drautz_2020,rinaldi,hoEquivariantManybodyMessage2026,zhengIntegratingDeepLearningBasedMagnetic2025a,xuSpininformedUniversalGraph2025a,gautamConvolutionalKernelEmbeddedE3Equivariant2026,wangCartesianTensorEquivariant2026,rayatGraphNeuralNetwork2026,uenoSpinMultiNetNeuralNetwork2024}.

We discuss all the intertwined atomic- and nanoscale degrees of freedom that need to be accurately described to obtain accurate magnon spectral functions and macroscopic spin-stiffness predictions, and we introduce a Coupled Optimization of SpIN Environment (COSINE) algorithm that allows us to achieve this goal by exploiting magnetic point-group symmetries. We demonstrate that COSINE converges faster than conventional simulated annealing approaches, and, most importantly, it also allows the optimization of the magnetic moment magnitudes.
After relaxing the atomic and magnetic-moment structure with atomic- and nanoscale nonhomogeneities, we rely on the tilted spins approach \cite{liechtenstein} to predict the exchange interactions with full sensitivity to these distortions and inhomogeneities, exploiting the finite receptive field of spin-informed machine-learned models to accelerate the calculations.
We then introduce paraunitary unfolding, a band-unfolding procedure for the paraunitary eigenvectors of bosonic Hamiltonians, from which we obtain the magnon spectral function.

We showcase our approach in a representative ferrimagnet for generality, and then combine it with a state-of-the-art SI-MLIP for iron \cite{rinaldi} to quantitatively predict how nonuniformities in the spatial distribution of vacancy defects control the blurring of the magnon spectral function. Finally, we discuss the implementation of our workflow in the \textit{Magnon} software package \cite{magnon-package}, which enables the use of our framework with any modern SI-MLIP.

\medskip
\noindent\textbf{Magnons in defective solids}\\
\nopagebreak
Magnons are often studied in solids that can be considered a perturbed version of an idealized system with long-range order \cite{buczekSpinWavesDisordered2018a, buczek_nonstoichiometric, jiang_feni_sro}. Formally, a periodic system has a finite Brillouin zone (BZ) where a magnon dispersion is well-defined \cite{kittel}; when defects such as vacancies \cite{beeler, mudrick}, dopants \cite{dilucente}, and interstitials \cite{chapman_mlip} are introduced, the long-range order is broken and, strictly speaking, the BZ reduces to the zero wavevector only---from a mathematical viewpoint, this implies that an ideal magnon bandstructure can no longer be defined.
However, from a physical viewpoint it is natural to ask how different the magnon spectral properties are between the ideal pristine and perturbed system.
To address this question, we introduce the theoretical and computational framework to quantitatively predict how the magnon spectral function is influenced by defects.

In general, the magnetic moments are sensitive to disorder in the structure and other degrees of freedom in a non-trivial way, and therefore it is desirable to rely on a procedure that does not require any prior assumption on their orientation and magnitude.
To this aim, we rely on the rotated-frame approach \cite{toth} to construct the most general Heisenberg Hamiltonian of a magnetic system with arbitrarily disordered atomic structure and composition.
Formally, this corresponds to a Bogoliubov--de~Gennes (BdG) bosonic Hamiltonian \cite{colpa}, which satisfies the most general paraunitary algebra.
Then, with the goal of constructing in full generality the spectral function from the eigenvalues of such a BdG Hamiltonian, we introduce a paraunitary unfolding procedure that can be used in the presence of paraunitary symmetries.
We will show how paraunitary unfolding generalizes the unfolding technique developed in Ref.~\cite{allen_unfolding} to determine the spectral function of Hamiltonians that describe defective systems and satisfy unitary algebra (e.g., electronic \cite{PhysRevB.89.041407, PhysRevB.91.041116} and phononic \cite{togo_unfolding}); in particular, the unitary case is obtained in the special limit of ferromagnetic collinear systems \cite{kittel}, and in the limiting case of vanishing defects the present approach recovers the pristine bandstructure.

We start by recalling the most general form of the Heisenberg Hamiltonian in linear spin-wave theory (LSWT) \cite{whiteQuantumTheoryMagnetism2007,toth,colpa},
\begin{equation}
\begin{split}
    H_S &= -\frac{1}{2} \sum_{\bm{r},\bm{r'}} \sum_{b,b'} \sum_{\alpha,\alpha'}  \hat{S}^\alpha_{\bm{r}b} J^{\alpha\alpha'}_{\bm{r}b\bm{r'}b'} \hat{S}^{\alpha'}_{\bm{r'}b'},
    \\
    &\approx \sum_{\bm{k}}
    \begin{pmatrix}
    \hat{b}_{\bm{k}b}^\dag & \hat{b}_{-\bm{k}b}
    \end{pmatrix}
    \begin{pmatrix}
        B_{1bb'}(\bm{k})&B_{2bb'}(\bm{k})\\
        (B_{2})^\dag_{bb'}(\bm{k})&\overline{B}_{1bb'}(-\bm{k})\\
    \end{pmatrix}
    \begin{pmatrix}
    \hat{b}_{\bm{k}b'} \\ \hat{b}_{-\bm{k}b'}^\dag
    \end{pmatrix},
\end{split}
\raisetag{20mm}
\label{eq:heisenberg_hamiltonian}
\end{equation}
where $\hat{S}^\alpha_{\bm{r}b}$ with $\alpha\in(x,y,z)$ is the spin angular momentum of atom $b$ (at position $\bm{\tau}_b$ in the unit cell) in the cell periodic image at $\bm{r}$, in units of $\hbar$, satisfying the commutation relations $[\hat{S}_{\bm{r}b}^{\alpha},\hat{S}_{\bm{r'}b'}^{\beta}] = i\epsilon_{\alpha\beta\gamma}\delta_{\bm{r}\bm{r'}}\delta_{bb'} \hat{S}_{\bm{r}b}^{\gamma}$.  
 
The tensor $J^{\alpha\alpha'}_{\bm{r}b\bm{r'}b'}$ is the exchange coupling, i.e., the curvature of the magnetic energy of the solid, which is quadratic around its minimum. The tensor interaction's on-diagonal components describe anisotropic direct exchange between atoms close enough to have wavefunction overlap, and indirect exchange between atoms that do not have significant direct wavefunction overlap, but do with the wavefunction of a common intermediary \cite{Coronado_1996} (these include RKKY from conduction electrons \cite{ruderman, kasuya, yosida}, superexchange \cite{anderson_superexchange} and Kitaev interactions \cite{Kitaev_2006, PhysRevLett.102.017205, rauGenericSpinModel2014}). 
The tensor's off-diagonal components describe the effects of spin--orbit coupling (SOC) through the Dzyaloshinskii--Moriya interaction \cite{Dzyaloshinsky_1958, moriya_anisotropic, PhysRevB.52.10239} and symmetric anisotropic exchange \cite{rauGenericSpinModel2014, shekhtman_dmi, moriya_anisotropic}.

The second line of Eq.~(\ref{eq:heisenberg_hamiltonian}) is the same Hamiltonian rewritten in the second-quantization formalism in Fourier space, as we show now.
First, we recall that the spin raising and lowering operators, $\hat{S}^{\pm}_{\bm{r}b} = \hat{S}^x_{\bm{r}b} \pm i \hat{S}_{\bm{r}b}^y$, can be recast via the Holstein--Primakoff transformation \cite{holstein} in terms of bosonic creation and annihilation operators for a spin-flip excitation in real space ($\hat{b}^\dagger_{\bm{r}b}$ and $\hat{b}_{\bm{r}b}$, respectively; these are related to the spin raising and lowering operators via $\hat{S}^+_{\bm{r}b} {=} \sqrt{2S_{\bm{r}b}}\left( 1-\frac{\hat{b}_{\bm{r}b}^{\dag}\hat{b}_{\bm{r}b}}{2S_{\bm{r}b}} \right)^{\frac{1}{2}}\hat{b}_{\bm{r}b} {\approx}\sqrt{2S_{\bm{r}b}}\hat{b}_{\bm{r}b}$).

We note, in passing, that while the operators $\hat{\bm{S}}^2_{\bm{r}b}$ and $\hat{S}^z_{\bm{r}b}$ have eigenvalues $S(S+1)$ and $m=-S,\ldots,S$, respectively, for integer or half-integer $S$, in an itinerant magnet, the spin magnitude $S_{\bm{r}b}$ inferred from the magnetic moment per atom is an average spin polarization of delocalized electrons, so it does not have to be integer or half-integer \cite{Skomski_2008, Stoner_1938}; additionally, it can also include contributions from the orbital angular momentum ($\lesssim 5\%$ in iron \cite{PhysRevLett.75.152}) and is sometimes used as an `effective' spin \cite{Abragam_1951}.
It is worth mentioning that sometimes the Heisenberg Hamiltonian is rewritten by absorbing the magnitude of the spin operators into the exchange tensor \cite{Evans_2014, PhysRevB.99.224414, Mentink_2010, Radu_2011}, and consequently using unit vector spins. These are two physically equivalent ways of writing the same Hamiltonian, related by a transformation discussed in the Methods \cite{chapman_nonheisenberg}; we reiterate that in this work we consider the convention in which the spin operators can have magnitude different from one.
Then, by applying a Fourier transform to these operators, one obtains the bosonic operators that create ($\hat{b}_{\bm{k}b}^\dag = \frac{1}{\sqrt{\mathcal{N}_{\rm c}}} \sum_{\bm{r}} \hat{b}_{\bm{r}b}^\dag e^{i\bm{k}\cdot(\bm{r}+\bm{\tau}_b)}$) or annihilate ($\hat{b}_{\bm{k}b}$) a magnon, the quantum of the spin wave, with wavevector $\bm{k}\in\mathcal{B}$, the first BZ of the crystal.
Finally, the expressions for the block matrices $B_{1bb'}(\bm{k})$ and $B_{2bb'}(\bm{k})$ are derived in the Methods using the rotated-frame approach; $\overline{B}_{1bb'}(\bm{k})$ is the complex conjugate of ${B}_{1bb'}(\bm{k})$, and 
$(B_2)^\dag_{bb'}(\bm{k})$ is the adjoint of $B_{2bb'}(\bm{k})$. 
We note, in passing, that in the most general case, the second line of Eq.~(\ref{eq:heisenberg_hamiltonian}) has the form of a matrix contracted with Nambu spinors \cite{tesfayeQuantumGeometryBosonic2025}; in the special simplified limit of an ideal ferromagnetic Hamiltonian \cite{kittel}, $B_{1bb'}(\bm{k}) = \overline{B}_{1bb'}(-\bm{k})$ and $B_{2bb'}(\bm{k})$ is zero in all elements (i.e., only the term proportional to the number operator ($\hat{b}_{\bm{k}b}^\dag\hat{b}_{\bm{k}b}$) survives), and it is sufficient to diagonalize $B_{1bb'}(\bm{k})$ only.

Eq.~(\ref{eq:heisenberg_hamiltonian}) applies to any magnetic system, ranging from pristine to disordered. In a pristine crystal, $\bm{r}$ runs over the $\mathcal{N}_{\rm c}$ Bravais-lattice sites and $\bm{k}$ over a corresponding set of $\mathcal{N}_{\rm c}$ wavevectors in the first BZ, and $b$ assumes values from one to the (small) number of atoms $N_{\rm at}$ in the unit cell. In the case of an infinite solid lacking translational symmetry, $\bm{r}$ and $\bm{k}$ can be discarded (both are restricted to zero only) and only the atomic index $b$ is used. 

From a physical viewpoint, within the LSWT considered here, creating an excitation of a spin configuration is equivalent to annihilating an excitation of the spin configuration in a reference frame rotated through $\pi$ radians.
For ideal ferromagnetic states, all the spin-wave excitations can be described in terms of the magnon creation operator $\hat{b}_{\bm{k}b}^\dag$, and the $B_{1bb'}(\bm{k})\hat{b}_{\bm{k}b}^\dag\hat{b}_{\bm{k}b'}$ terms in Eq.~(\ref{eq:heisenberg_hamiltonian}) conserve their number.
In contrast, in collinear antiferromagnets, a magnon creation operator can combine $\hat{b}_{\bm{k}b}^\dag$ on the up sublattice with $\hat{b}_{-\bm{k}b'}$ on the down sublattice (correspondingly, the terms $B_{2bb'}(\bm{k}) \hat{b}_{\bm{k}b}^\dag \hat{b}_{-\bm{k}b'}^\dag$ and $(B_{2})^\dag_{bb'}(\bm{k}) \hat{b}_{-\bm{k}b} \hat{b}_{\bm{k}b'}$ create and annihilate pairs of local spin deviations in Eq.~(\ref{eq:heisenberg_hamiltonian})).
The particle--hole symmetry \cite{tesfayeQuantumGeometryBosonic2025, harmsAntimagnonics2024a} relates the $2N_{\rm at}$ eigenvalues of the doubled eigenproblem in pairs at opposite wavevectors, giving $N_{\rm at}$ independent magnon branches.

This particle--hole symmetry is also reflected in the structure of the eigenvectors $\mathcal{T}_{\bm{k}b\eta}$ of the magnon Hamiltonian in Eq.~(\ref{eq:heisenberg_hamiltonian}), where the index $\eta$ differentiates the $N_{\rm at}$ particle ($\eta=+1$, corresponding to creation operators) elements from the $N_{\rm at}$ hole ($\eta=-1$, corresponding to annihilation operators) elements. Preserving the bosonic commutators requires the matrix of these eigenvectors to satisfy \cite{colpa}
\begin{equation} \label{eq:paraunitary_condition}
\sum_{b_1\eta_1}\sum_{b_2\eta_2}\mathcal{T}_{\bm{k},b\eta,b_1\eta_1}\Phi_{b_1\eta_1,b_2\eta_2}(\mathcal{T}^\dag)_{\bm{k},b_2\eta_2,b'\eta'}=\Phi_{b\eta,b'\eta'},
\end{equation}
where the para-identity matrix $\Phi=\operatorname{diag}(I_{N_{\rm at}}, -I_{N_{\rm at}})$ is a block-diagonal matrix and $I_{N_{\rm at}}$ is the $N_{\rm at}\times N_{\rm at}$ identity matrix \cite{colpa}. Equivalently, $\Phi=\sigma_z\otimes I_{N_{\rm at}}$ for the Pauli spin matrix $\sigma_z$ \cite{tesfayeQuantumGeometryBosonic2025}. Henceforth we will refer to eigenvectors satisfying Eq.~(\ref{eq:paraunitary_condition}) as ``paraunitary eigenvectors''.

In the next section, we perform the paraunitary diagonalization of the magnon Hamiltonian following Ref.~\cite{colpa} (see Methods). First, we determine the magnon bandstructure for a pristine system, and then leverage our paraunitary unfolding framework to investigate how such bandstructure (the non-interacting spectral function) is modified by defects.

\section*{RESULTS}

\medskip
\noindent\textbf{Paraunitary unfolding \& spectral function}\\
\nopagebreak
In the previous section we introduced the general framework to describe systems ranging from pristine crystals to completely disordered systems.
In a material containing defects, structural imperfections, or other types of disorder, the Hamiltonian is expected to approximately exhibit the translational symmetry of the parent lattice \cite{allen_unfolding};
it is therefore instructive to ``unfold'' the corresponding eigenvalues in the BZ of the parent crystal, yielding a spectral function which, upon increasing disorder, becomes an increasingly blurred (and possibly shifted) version of the pristine bandstructure \cite{buczekSpinWavesDisordered2018a}.

In this section we discuss how to construct such a spectral function representation in systems with arbitrary magnetic configuration (described by the full paraunitary Hamiltonian in Eq.~(\ref{eq:heisenberg_hamiltonian})) and therefore applicable to magnon excitations in antiferromagnets, ferrimagnets, and magnetic materials with non-collinearities.
We will show how our procedure generalizes past approaches for phonons \cite{baroni_unfold,togo_unfolding,allen_unfolding}, electrons \cite{popescu,allen_unfolding,PhysRevLett.80.4725}, and ferromagnets \cite{buczekSpinWavesDisordered2018a} limited to Hamiltonians with unitary symmetries.   

To illustrate the procedure, we recall that to model defects such as vacancies and compositional substitutions in a crystal, the established explicit (beyond-mean-field) approach usually builds a large supercell of the pristine system, and then introduces vacancy defects into it \cite{chapman_mlip, PhysRevB.106.L060403, mudrick}.  
To model such a defective system, we start by discussing the mapping between equivalent descriptions of the Hamiltonians (i.e., eigenvectors and eigenvalues) of a crystal in the primitive cell (`pc') and supercell (`sc') representations, and then discuss how such a mapping generalizes when defects are introduced in the sc representation.

\begin{figure}[b]
    \centering
    \includegraphics[width=\columnwidth]{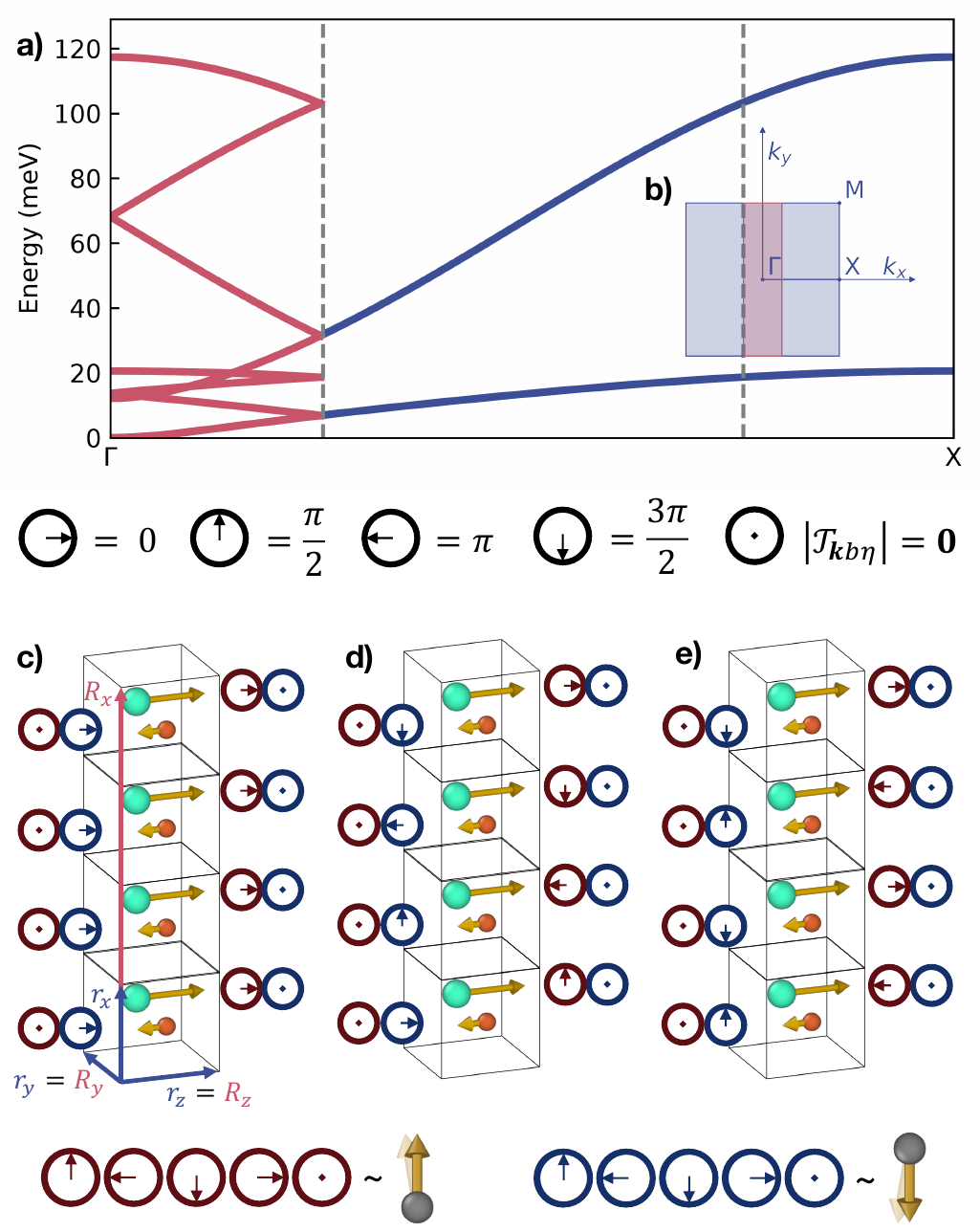}
    \caption{
        (a) Paraunitary unfolding of the magnon bands of a model ferrimagnet from the $4{\times}1{\times}1$ supercell (red) to the primitive cell (blue). (b) The BZs of the two cells, whose volume is inversely proportional to the volume of the real-space unit cells. The model ferrimagnet has a CsCl structure with antiparallel spins $S_{\rm A}=3/2$ and $S_{\rm B}=7/2$, coupled by nearest-neighbor exchange interactions $J_{\rm A\text{-}A}=18$~meV, $J_{\rm B\text{-}B}=5$~meV and $J_{\rm A\text{-}B}=-2$~meV.
        (c--e) The $i\bm{G}\cdot\bm{r}$ complex phases of the paraunitary eigenvectors of the $4{\times}1{\times}1$ supercell; the three eigenvectors correspond to the three segments of the $\Gamma$--$X$ path in (a). Phases are represented by unit vector Argand diagrams and indicated by a red (blue) clock for the `particle' $\eta{=}{+}1$ (`hole' $\eta{=}{-}{1}$) parts of the eigenvector.
        The relationship between phases derives from the translational symmetry of the ideal lattice. 
    }
    \label{fig:unfolding}
\end{figure}

We denote with $\bm{r}\in \mathcal{R}^{\rm pc}$ the direct lattice vectors of a perfect crystal in the pc representation, and correspondingly wavevectors $\bm{k}\in \mathcal{B}^{\rm pc}$. Similarly, for the sc representation, we denote the corresponding quantities with $\bm{R}\in \mathcal{R}^{\rm sc}$ and $\bm{K}\in\mathcal{B}^{\rm sc}$. 
These quantities are identified in Fig.~\ref{fig:unfolding}, which gives two equivalent example representations: the pc and a sc that is a $4\times 1 \times 1$ repetition of the pc.
Formally, the equivalence between pc and sc representations can be understood from the relation between the paraunitary eigenvectors in these representations, specifically from the behavior of the sc eigenvectors under pc translations, as shown in Fig.~\ref{fig:unfolding}\textbf{c--e}. 
Specifically, we see that in the sc eigenvectors ${\mathcal{T}}^{\rm sc}_{\bm{k}b\eta}$ both the particle ($\eta=+1$) and hole ($\eta=-1$) sectors have a constant phase difference between elements related by the primitive lattice translation operator $\hat{T}_{\bm{r}}$ (its action is to rigidly shift atoms in real space by a primitive lattice vector $\bm{r}$). This constant phase difference stems from the fact that $\hat{T}_{\bm{r}}$ acts in the particle and hole subspaces independently (i.e., without mixing them); mathematically, its independent action in these subspaces corresponds to a permutation in the map that relates the eigenvector's atomic index $b$ to the atom in the supercell: 
\begin{equation} \label{eq:permutations}
    \hat{T}_{\bm{r}}{\mathcal{T}}^{\rm sc}_{\bm{k}b\eta} = {\mathcal{T}}^{\rm sc}_{\bm{k}\:[\hat{T}_{\bm{r}}]^{-1}(b)\:\eta}.
\end{equation}
Since the eigenvectors of the primitive cell Hamiltonian satisfy Bloch's theorem and are thus modulated plane waves, and a real-space translation of a plane wave corresponds to a phase shift, it is instructive to interpret the action of $\hat{T}_{\bm{r}}$ in reciprocal space.
In particular, a wavevector $\bm{k}$ bounded within the large BZ of the primitive crystal can be decomposed into the sum $\bm{k}=\bm{G}+\bm{K}$, of a supercell reciprocal lattice vector $\bm{G}$ and a remainder $\bm{K}$.
As shown in Figs.~\ref{fig:unfolding}\textbf{a,b}, within equivalent pc and sc representations of the same perfect crystal, one can directly unfold bands from $\mathcal{B}^{\rm sc}$ into $\mathcal{B}^{\rm pc}$, since the task effectively reduces to repeatedly ``mirroring'' the sc bands (formally, this is done by identifying which $\bm{G}$ each folded band in the sc representation has to be unfolded to).
In this ideal crystal case, the single $\bm{G}$ vector to which each sc band corresponds is unambiguously identified by the property $\hat{T}_{\bm{r}}\mathcal{T}^{\rm sc}_{\bm{k}b\eta} = e^{i\bm{G}\cdot\bm{r}}\mathcal{T}^{\rm sc}_{\bm{k}b\eta}$.

In the general case, when a material has disorder and so respects translational symmetry only approximately, a supercell eigenvector has non-negligible projections over several possible pc Bloch states, each modulated by a specific $\bm{G}$ in different sc BZs. This is captured by unfolding weights 
\begin{equation} \label{eq:unfolding_weight}
    W_{\bm{K}b\eta}(\bm{G}) = \frac{1}{\mathcal{N}_{\rm c}^2} \frac{ \left \lVert \sum_{\bm{r}} e^{-i\bm{G}\cdot\bm{r}}\hat{T}_{\bm{r}} \mathcal{T}^{\rm sc}_{\bm{k}b\eta} \right \rVert^2}{\left \lVert \mathcal{T}^{\rm sc}_{\bm{k}b\eta} \right \rVert^2},
\end{equation}
which use $\hat{T}_{\bm{r}}$ to find the Fourier coefficient of each $\bm{G}$ component. 
The interpretation of the paraunitary unfolding weight is as the $\bm{G}$ Fourier component of the magnon eigenvectors---in both the particle and hole sectors---normalized by the (square of the) Euclidean norm of the paraunitary eigenvectors $\lVert \mathcal{T}^{\rm sc}_{\bm{k}b\eta} \rVert^2 = (\mathcal{T}^{\rm sc}_{\bm{k}b\eta})^\dag \mathcal{T}^{\rm sc}_{\bm{k}b\eta}$. 
From the paraunitary property in Eq.~(\ref{eq:paraunitary_condition}), which defines the (squared) ``paranorm'' $\lVert \mathcal{T}^{\rm sc}_{\bm{k}b\eta} \rVert_\mathcal{P}^2 = \lVert \mathcal{T}^{+}_{\bm{k}b\eta} \rVert^2 - \lVert \mathcal{T}^{-}_{\bm{k}b\eta} \rVert^2 = \pm1$, where $\mathcal{T}^{+}_{\bm{k}b\eta}$ and $\mathcal{T}^{-}_{\bm{k}b\eta}$ are the particle and hole components of $\mathcal{T}^{\rm sc}_{\bm{k}b\eta}$, it can be shown that the (squared) Euclidean norm $\lVert \mathcal{T}^{\rm sc}_{\bm{k}b\eta} \rVert^2 = \lVert \mathcal{T}^{+}_{\bm{k}b\eta} \rVert^2 + \lVert \mathcal{T}^{-}_{\bm{k}b\eta} \rVert^2 \geq 1$, ensuring that the unfolding weight is well defined.
In the special case of a ferromagnet with $B_{2bb'}(\bm{k})$ zero and Hamiltonian proportional to $\hat{b}_{\bm{k}b}^\dag\hat{b}_{\bm{k}b}$, the Hamiltonian has unitary eigenvectors with $\lVert \mathcal{T}^{\rm sc}_{\bm{k}b\eta} \rVert = 1$, in which case Eq.~(\ref{eq:unfolding_weight}) 
is analogous to the electron or phonon unfolding discussed in Refs.~\cite{allen_unfolding,togo_unfolding,Mayo2020Unfolding,dilucente}.

For later convenience, it is useful to mention that the weights defined in Eq.~(\ref{eq:unfolding_weight}) satisfy normalization and bound properties: the identity $\sum_{\bm{G}} e^{-i\bm{G}\cdot\bm{r}} = \mathcal{N}_{\rm c}\delta_{\bm{r},\bm{0}}$ implies that $\sum_{\bm{G}} W_{\bm{K}b\eta}(\bm{G}) = 1$ for any given $(\bm{K},b,\eta)$; the triangle inequality implies $0\le W_{\bm{K}b\eta}(\bm{G}) \le 1$.

\begin{figure*}[!ht]
    \centering
    \includegraphics[width=\textwidth]{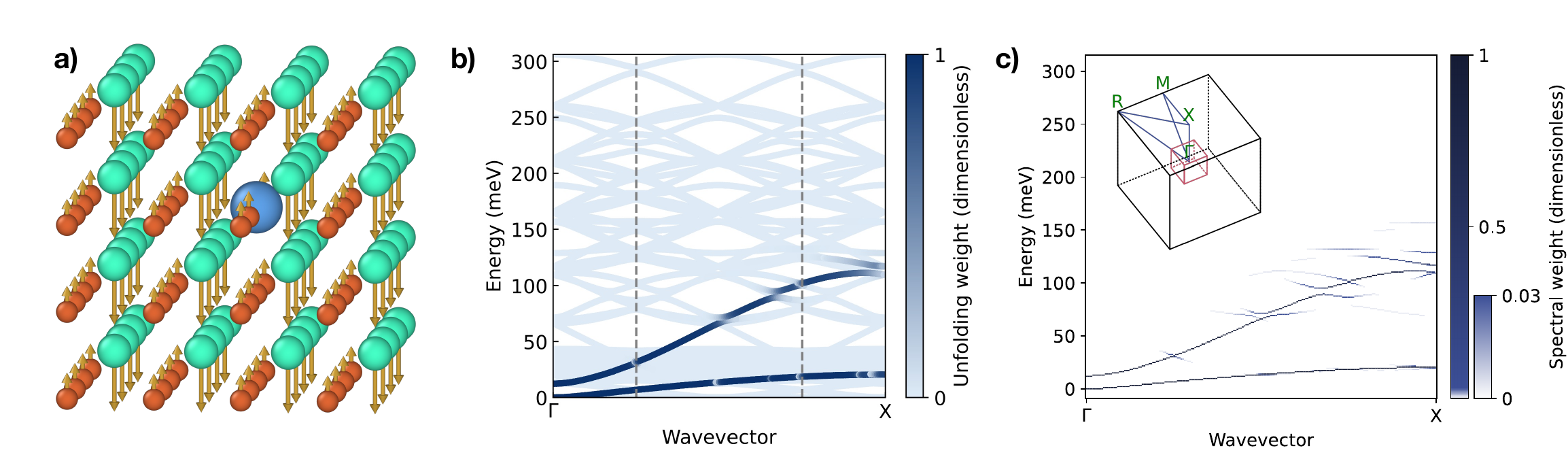}
    \caption{(a) A $4\times4\times4$ supercell of the fictitious ferrimagnetic model with a single A site replaced by a nonmagnetic substitution defect (blue sphere). (b) The unfolded bands computed using Eq.~(\ref{eq:unfolding_weight}). The breaking of translational symmetry means that the weights now exist in a continuum between 0 and 1, in contrast to the binary weights of the perfect supercells in Fig.~\ref{fig:unfolding}\textbf{a}. (c) The spectral function obtained by weighted binning of the unfolded bands. Inset, the BZ and high-symmetry points of the unit cell (black) and supercell (red) for this system.}
    \label{fig:unfolding2}
\end{figure*}

The weights in Eq.~(\ref{eq:unfolding_weight}) are binary for translation eigenstates in the ideal-crystal limit; as disorder increases, they become more broadly distributed over the interval $[0,1]$.
They represent how a single modulated plane-wave eigenstate of the pc in the perfect crystal is redistributed to multiple such states in the disordered system. 
Therefore, to see how different the spectrum of the disordered sc is compared to that of the pristine pc, we use the paraunitary unfolding weights to determine the spectral function \cite{allen_unfolding}
\begin{equation} \label{eq:spectral_function_from_unfolding}
    A(\bm{k}{=}\bm{G}{+}\bm{K}, E) = \sum_{b,\eta} W_{\bm{K}b\eta}(\bm{G}) \delta(E-E_{b\eta}(\bm{G}{+}\bm{K})).
\end{equation}
For the physical positive-energy spectrum, the sums over $(b,\eta)$ here and below run over the independent positive-norm modes, with both particle and hole components included in each weight. The aforementioned normalization and bound properties of the weights imply that the magnon density of states obtained by integrating the disorder-blurred spectral function $A(\bm{k}, E)$ in $\mathcal{B}^{\rm pc}$ ($G^{\rm pc}(E)$) is equal to that obtained by integrating the non-interacting (Dirac-delta) spectral function (i.e., the perfect bands) in $\mathcal{B}^{\rm sc}$ ($G^{\rm sc}(E)$):
\begin{align}
    G^{\rm pc}(E) &=  \!\!\!\!\sum_{b,\eta,\bm{k}=\bm{G}{+}\bm{K}\in \mathcal{B}^{\rm pc}} \!\!\!\!W_{\bm{K}b\eta}(\bm{G}) \delta(E{-}E_{b\eta}(\bm{G} {+} \bm{K}))\\
    &= \sum_{b,\eta, \bm{K}\in \mathcal{B}^{\rm sc}} \delta(E-E_{b\eta}(\bm{K})) =G^{\rm sc}(E),
\end{align}
where we used the periodicity property $E_{b\eta}(\bm{k}) = E_{b\eta}(\bm{G} + \bm{K}) = E_{b\eta}(\bm{K})$. 

In Fig.~\ref{fig:unfolding2} we showcase the procedure with a model ferrimagnet with a single vacancy defect (panel \textbf{a}), highlighting how the unfolded bands (panel \textbf{b}) are used to compute the spectral function using Eq.~(\ref{eq:spectral_function_from_unfolding}) (panel \textbf{c}). We stress that the broadening of the spectral function is due to defects only, reflecting the capability of this approach to isolate the contribution of disorder from other intrinsic effects.
Because these broadening effects are largely temperature independent, in the low-temperature limit they are expected to dominate over intrinsic, temperature-activated effects such as magnon--magnon scattering \cite{zhitomirsky_magnon_decays, luscher_exact_diagonalization}. 
As we will see, the procedure introduced here could be useful to assess the concentration and spatial distribution of defects from experimental measurements of magnon spectra. 

We remark that using the paranorm $\lVert \mathcal{T}^{\rm sc}_{\bm{k}b\eta} \rVert_\mathcal{P}^2 = (\mathcal{T}^{\rm sc}_{\bm{k}b\eta})^\dag \Phi \mathcal{T}^{\rm sc}_{\bm{k}b\eta}$ in place of the Euclidean norm gives the Krein signature of the projection, whose sign captures the particle or hole character of the unfolded eigenvectors \cite{tesfayeQuantumGeometryBosonic2025, massarelli_krein}. The sum over the resulting weights at all $\bm{G}$ is always one because a particle must unfold onto a particle (and a hole onto a hole), but in the presence of disorder supercell eigenvectors may unfold onto states in the primitive cell with opposite Krein signature, so individual weights can be negative with this choice of norm.

\medskip
\noindent\textbf{COSINE spin-lattice energy minimization}\\
\nopagebreak
Because magnons are perturbations of the spin degrees of freedom relative to a ground-state configuration, to simulate them one has to find a minimum of the magnetic potential energy surface (mPES), and then compute how the material responds to perturbations of the spin degrees of freedom around this mPES minimum.
In real materials with defects, finding such a minimum of the mPES is more challenging compared to the illustrative example shown in the previous section.
In this section, we introduce the Coupled Optimization of SpIN Environment (COSINE) algorithm to efficiently and robustly find local mPES minima with stable magnons, and discuss how this method enables us to simulate atomically disordered magnets with non-trivial spatial correlations at the nanoscale.

Minimization methods need to simultaneously address the challenges of capturing: (i) the coupling between spin and structural degrees of freedom \cite{tranchida}, (ii) longitudinal spin fluctuations (magnetic moment variations) \cite{dudarev_longitudinal, chapman_nonheisenberg}, and (iii) interactions that decay over distances larger than the typical size of DFT simulations \cite{pajda, marzari_electronic_structure}.

The three challenges have traditionally been mitigated by fixing the magnitudes of the magnetic moments to be constant \cite{tranchida, evansAtomisticSpinDynamics2018b} or by employing semi-empirical parameterizations \cite{mudrick,buczekSpinWavesDisordered2018a}, especially to describe the coupling between spin magnitude, exchange and atomic structure.  

Recently, the development of SI-MLIPs \cite{rinaldi,chapman_mlip,hoEquivariantManybodyMessage2026,wangCartesianTensorEquivariant2026, yangDeepLearningIlluminates2024,dengCHGNetPretrainedUniversal2023a,yuSpindependentGraphNeural2024,eckhoffHighdimensionalNeuralNetwork2021,novikovMagneticMomentTensor2022,xuSpininformedUniversalGraph2025a,wuLocalenergydrivenMachinelearningExchange2026,shiAtomisticPredictionMagnetic2026,rayatGraphNeuralNetwork2026} has made it possible to address challenge (iii) without significantly losing accuracy; however, challenges (i) and (ii) are still difficult because of the practical finding that, as we will see, the mPES can be rough with many saddle points.

To address these challenges and robustly determine the minimum of an mPES without any assumption on its smoothness, we introduce the COSINE algorithm. Such an algorithm leverages physical symmetries to guide the search, and uses energies rather than forces, since the latter are more affected by noise and can be strongly oscillatory in the presence of a rough mPES.

We begin by outlining the two key observations that motivate the design of the algorithm. First, in the case of dilute, homogeneous defects it is foreseeable that the majority of defects will be well separated. This means that, across a large system, there are likely to be many atoms with similar LAEs; therefore, within a Metropolis--Hastings Monte Carlo \cite{hastingsMonteCarloSampling1970} (MC) optimization,
it would be computationally economical to share successful 
moves between these atoms. Second, two atoms around a defect cluster will have similar energies when their positions relative to the cluster are related by magnetic point-group symmetry operations of the crystal. This is most easily seen for a single point defect where the atoms in each interaction shell around it have the same LAE. We expect that grouping atoms with symmetry-equivalent LAEs and performing collective MC \cite{swendsenNonuniversalCriticalDynamics1987,wolffCollectiveMonteCarlo1989}
trial moves synchronized across multiple atoms will improve the acceptance probability if the grouped atoms have significantly overlapping LAEs; the intuitive rationale behind this is that the equivalent chain of per-atom moves is unlikely, and the individual moves may not be energetically favorable. 

\begin{figure}[t]
    \centering
    \includegraphics[width=\columnwidth]{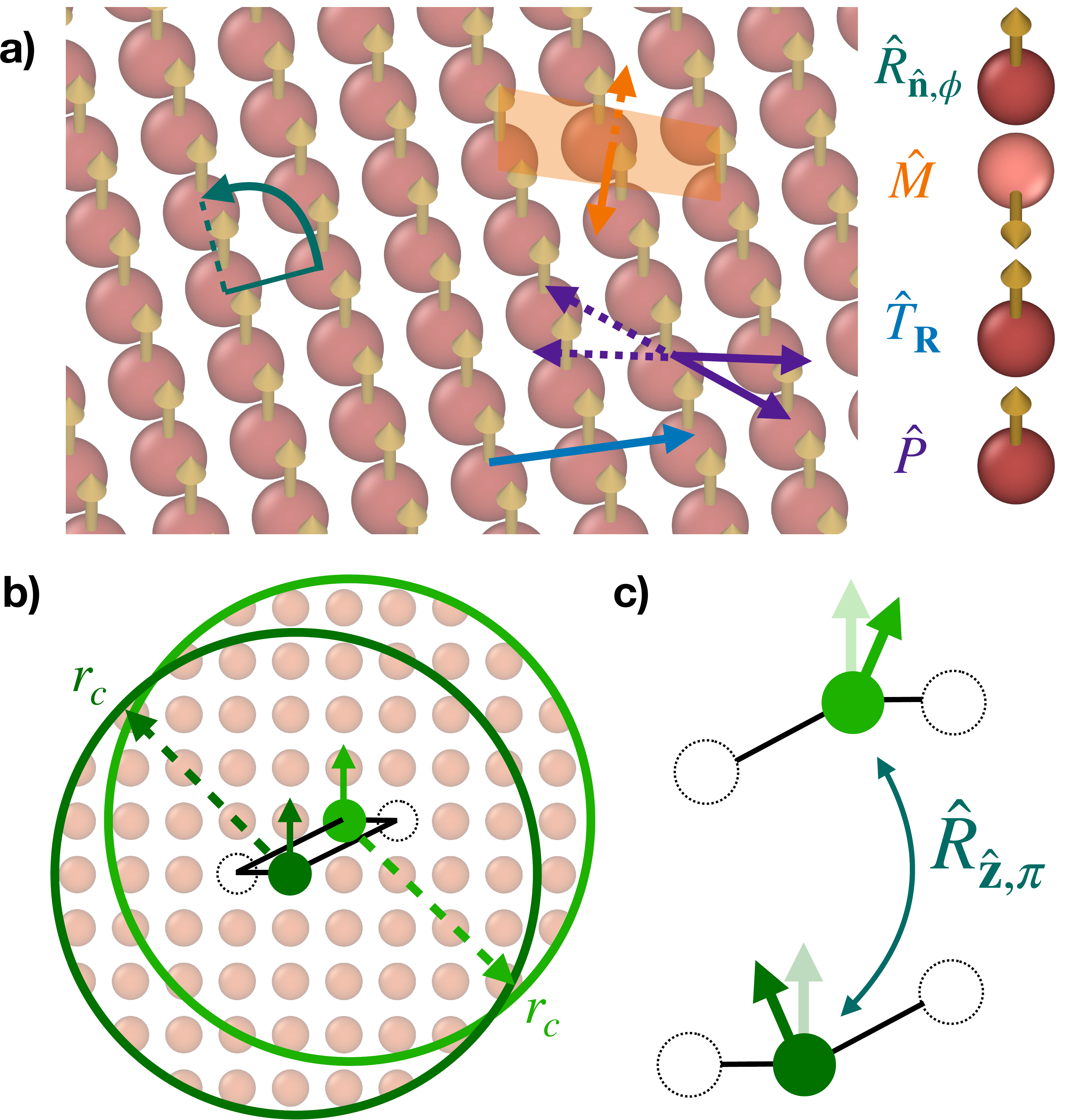}
    \caption{Schematic of the COSINE algorithm. (a) The symmetry operations of the lattice and their effect on the magnetic moment: rotations, mirrors and inversion, which form the magnetic point group used by COSINE, and translations, which leave the moment unchanged. (b) Simplified representation of the atomic clusters showing two cluster-equivalent sites based on their positions relative to a vacancy and related by rotational symmetry. (c) Example of a COSINE MC move applied to the magnetic moments, mapped between the clusters by application of the same symmetry relationship satisfied by those clusters.}
    \label{fig:relaxation}
\end{figure}

We now explain how our COSINE algorithm synchronizes moves across atoms with symmetry-equivalent LAEs and simultaneously relaxes the atomic, magnetic and cell degrees of freedom. 
The LAE of a magnetic atom is taken as the cluster $\bm{\sigma}$ of the surrounding $N_\text{LAE}$ atoms within a radial cutoff $r_c$, formally the set of position vectors $\bm{\tilde{r}}_i$, spin vectors $\bm{S}_i$, and in the case of a multicomponent system, atomic numbers $Z_i$,
\begin{equation}
    \bm{\sigma} = \{ \bm{\tilde{r}}_i, \bm{S}_i, Z_i \mid 0 \le i<N_\text{LAE} \}.
\end{equation}
Two atoms with clusters $\bm{\sigma}$, $\bm{\sigma'}$ then have their moves synchronized if they are equivalent under a magnetic point-group symmetry operation $\hat{O}$ \cite{spglib} of the pristine crystal, $\bm{\sigma} \equiv \hat{O}\bm{\sigma'}$, where
\begin{equation}
    \hat{O}\bm{\sigma'} = \{ \hat{O}\bm{\tilde{r}'}_i, \zeta_{\hat{O}}\det(\hat{O})\hat{O}\bm{S'}_i, Z'_i \mid 0\le i<N_\text{LAE} \}.
\end{equation}
The factor $\det(\hat{O})$ appears because the magnetic moment is a pseudovector \cite{Perez_Mato_2015, Shinohara_2023}, and $\zeta_{\hat{O}}=\pm1$ accounts for possible time-reversal symmetry with each $\hat{O}$. 
We note that this transformation is generally applicable even to systems with SOC, because it transforms the positional and magnetic degrees of freedom together. In the special case of negligible SOC, there is an additional symmetry under a uniform rotation of all spins, because their reference frame is no longer `locked' to the lattice by SOC. 
The procedure is outlined in Fig.~\ref{fig:relaxation}; using the symmetries of the crystal (panel \textbf{a}), the two atoms with equivalent LAEs are grouped together (panel \textbf{b}). When the magnetic moment of one of the atoms is updated, the update is synchronized across all other atoms in the group using the symmetry relationships between their clusters (panel \textbf{c}).

The COSINE algorithm may be applied when the interactions decay with distance sufficiently quickly that it is suitable to approximate them as zero beyond the cluster cutoff $r_c$ \cite{thomas_coulomb}. From a practical viewpoint, the identification of symmetry-equivalent LAEs is exact when employed with short-range MLIPs such as spin-informed ACE \cite{drautz_2020,rinaldi} using a cluster cutoff equal to the MLIP cutoff. More generally, the cluster size in COSINE does not need to be the same as the MLIP receptive field; using a smaller cutoff in COSINE will tend to overestimate degeneracies between atoms, since a smaller local environment provides less information to discriminate between them. Using a larger cutoff would slow down the calculation without any benefit, by splitting groups of atoms which are degenerate under the MLIP into smaller subgroups when they could be synchronized without any loss of generality. We note, in passing, that the current application of COSINE focuses on perturbative disorder, where the magnetic point-group symmetries of LAEs in the neighborhood of a vacancy can be related; more generally, in strongly disordered systems it may be more challenging to find relations between magnetic point-group symmetries of LAEs, and this regime will be subject of future work.

\begin{figure}[b]
    \centering
    \includegraphics[width=\columnwidth]{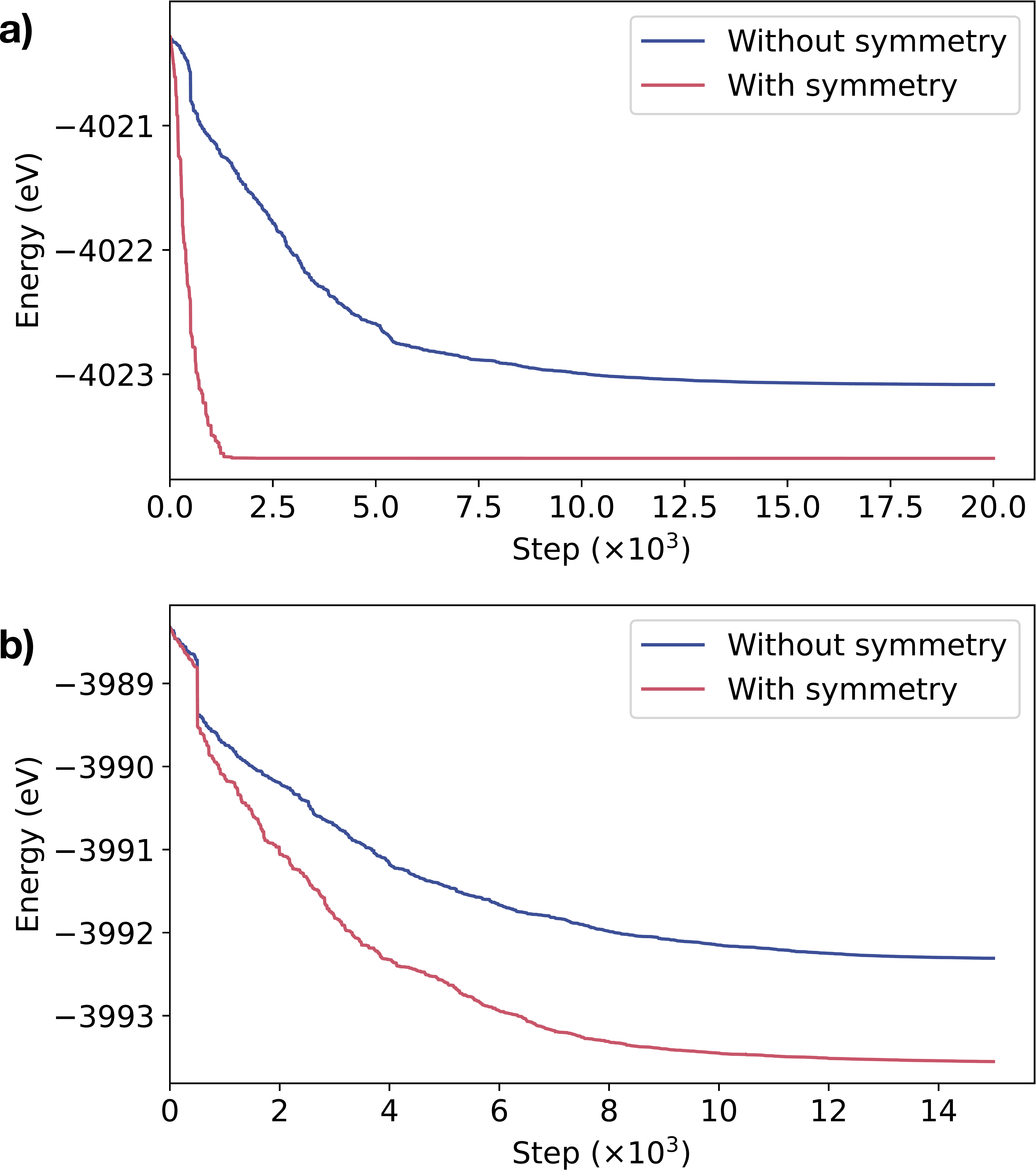}
    \caption{The energy trajectories from optimizations of two demonstrative 512-atom systems with (a) an isolated vacancy and (b) an asymmetric 4-vacancy cluster, benchmarking the COSINE algorithm against a conventional per-atom MC approach.}
    \label{fig:CECMC_convergence}
\end{figure}

Fig.~\ref{fig:CECMC_convergence} benchmarks the COSINE algorithm against a conventional (not collective) MC algorithm in a system with an isolated monovacancy, and a system with an asymmetric 4-vacancy cluster.  
The rate of convergence is increased by a factor of around seven in Fig.~\ref{fig:CECMC_convergence}\textbf{a} when moves are shared between atoms with symmetry-related LAEs, and the energy drop from the common starting configuration is about $30\%$ larger with COSINE than with standard per-atom optimization (Fig.~\ref{fig:CECMC_convergence}\textbf{b}). When considering real systems, the algorithm was used to optimize the same structural configuration multiple times but with different initial magnetic moments to verify that the relaxed structure is independent of the starting configuration. The scaling of computational cost depends only on the number of different LAEs under the magnetic point-group symmetries (which in turn is determined by the defect concentration and $r_c$).

\medskip
\noindent\textbf{Spatially correlated defects \& spin stiffness} \\
\nopagebreak
The framework introduced in the previous sections, capable of simulating magnons in systems with local compositional, structural, and magnetic-moment inhomogeneities, is especially useful to investigate real systems with non-trivial distributions of defects.

Iron is a paradigmatic ferromagnet that can feature vacancy defects with non-trivial spatial correlations due to manufacturing processes \cite{liang_wustite, nanda_3rd_gen_ahss}, radiation-induced damage \cite{beeler}, deformation \cite{yang_microvoid} and presence of hydrogen impurities \cite{sugita}. 
To investigate how non-uniformly distributed defects influence magnons, we start by classifying their isotropic but inhomogeneous distribution using a magnetic-defect correlation function
\begin{equation} \label{eq:correlator}
    C(\tilde{r}) {=} \frac{1}{N_{\rm d} \rho\, 4\pi \tilde{r}^2} \left\langle \sum_{\bm{r}bb'}\!\!\! \Delta_{\bm{r}b}\Delta_{\bm{0}b'}\,\delta\big(\tilde{r}{-}|\bm{\tilde{r}}_{\bm{0}b'}{-}\bm{\tilde{r}}_{\bm{r}b}|\big) \!\!\right\rangle,
\end{equation}
which is the magnetic counterpart to the mass disorder correlation function used in previous phonon studies \cite{thebaud,fiorentino}. Here, $\Delta_{\bm{r}b} \in \{0,1\}$ describes the discrete occupation by point defects of site $b$ in the periodic image $\bm{r}$ of the simulation cell, whose position is $\bm{\tilde{r}}_{\bm{r}b}=\bm{r}+\bm{\tau}_b$; $\delta(\tilde{r})$ is the Dirac delta at distance $\tilde{r}=|\bm{\tilde{r}}|$, and there are $N_{\rm d}$ defects with average number density $\rho$. The function $C(\tilde{r})$ can therefore be seen as the radial distribution function of magnetic point defects, since $\Delta_{\bm{r}b}\Delta_{\bm{0}b'}$ is nonzero only when both sites are occupied by defects. Intuitively, it is the factor by which the average number of defects at distance $\tilde{r}$ from any other defect differs from the homogeneous-distribution limit (in this homogeneous limit, for $N_{\rm at}\to\infty$, the sum can be replaced by a volume integral over a uniform density, yielding $C(\tilde{r})=1$). When the defects are not homogeneous, $C(\tilde{r})$ can decay in space, in our case as a power law of $\tilde{r}$ back to the baseline $C(\tilde{r})=1$, because at larger $\tilde{r}$ each spherical shell samples more sites, so local inhomogeneities make a progressively smaller contribution. The deviations of $C(\tilde{r})$ from one allow us to distinguish systems with different spatial correlations at the nanoscale. 

\begin{figure*}[!ht]
    \centering
    \includegraphics[width=\textwidth]{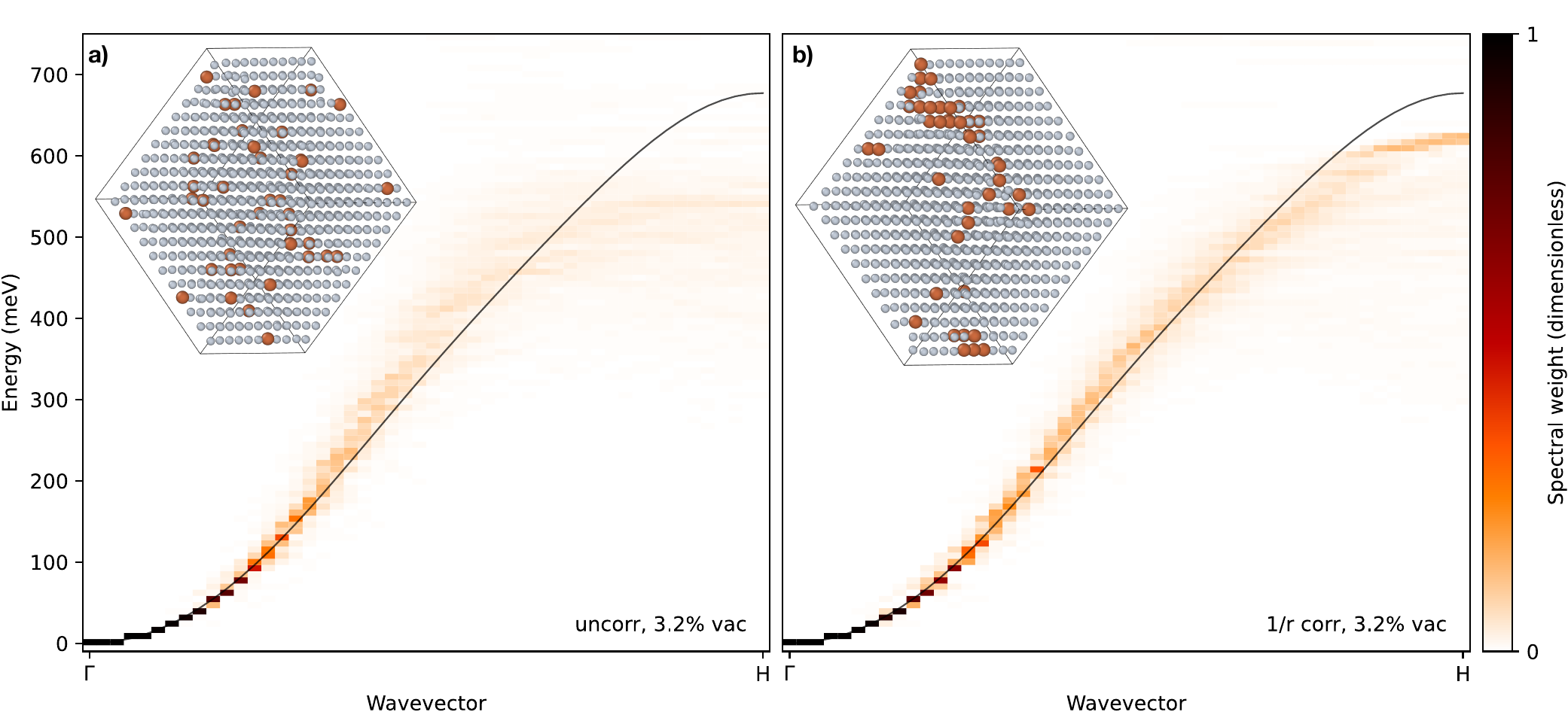}
    \caption{The magnon spectral function of 1331-atom systems of iron with (a) 3.2\% concentration of homogeneous vacancies and (b) 3.2\% concentration of vacancies distributed with $C(\tilde{r})\sim1/\tilde{r}$. The inset structures show the systems with iron sites colored silver and vacancy sites colored orange, and the black line is the Dirac delta spectral function of pristine iron. The blurring is solely due to disorder.}
    \label{fig:spectral function}
\end{figure*}

We investigate the effects of spatial correlations by comparing the magnon spectral function in the presence of uniformly distributed vacancies ($C(\tilde{r})=1$) and with spatial correlations $C(\tilde{r})\sim1/\tilde{r}$. 
Since iron is an itinerant magnet, the exchange interactions can be dependent on magnetic moment orientations and magnitudes \cite{kvashninMicroscopicOriginHeisenberg2016,szilvaTheoryNoncollinearInteractions2017}, and we compute them using the small-spin-tilts method \cite{liechtenstein, Rezaei_2026, RevModPhys.95.035004}:
\begin{equation}
    \delta E_{ij}(\theta) - \delta E_{i}(\theta) - \delta E_{j}(\theta) \approx \frac{1}{2}J_{ij}\theta^2,
    \label{eq:liechtenstein}
\end{equation}
where $\delta E_i(\theta)$ and $\delta E_j(\theta)$ are the energy changes induced by tilting by the angle $\theta$ spin $i$ and $-\theta$ spin $j$, respectively; and $\delta E_{ij}(\theta)$ is the energy cost of simultaneously tilting spin $i$ by $\theta$ and spin $j$ by $-\theta$ (see Fig.~\ref{fig:liechtenstein} in the Methods for details).
Determining all the exchange parameters involves a number of energy calculations of the order of $N_{\rm at}^2$,
and because simulating nanoscale correlations requires $N_{\rm at}\gtrsim 10^3$ atoms, these calculations are prohibitively challenging for DFT. 

We overcome this challenge by using the magnetic ACE SI-MLIP for iron from Ref.~\cite{rinaldi}, which was trained on defective structures, through the LAMMPS SPIN package \cite{lammps,tranchida}.
To capture the effects on the exchange interactions of atomic and nanoscale inhomogeneities in both the structural and magnetic degrees of freedom, and their combined effects on the magnon spectral function, we generated $1331$-atom ($11^3$) body-centered cubic (BCC) iron cells with a $3.2\%$ vacancy concentration, either uncorrelated or $\sim1/\tilde{r}$-correlated.
Because of the local architecture of this SI-MLIP, we evaluate Eq.~(\ref{eq:liechtenstein}) by accounting only for the LAE within a distance from the tilts defined by the finite receptive field of the SI-MLIP \cite{thomas_coulomb}, implying a computational cost linear in $N_{\rm at}$.

Overall, these developments enable us to simulate the magnon spectrum of the Hamiltonian in Eq.~(\ref{eq:heisenberg_hamiltonian}) accounting for the deviations from perfect ferromagnetism induced by disorder with different spatial correlations \cite{PhysRevB.38.11718,chapman_mlip}. Fig.~\ref{fig:spectral function}\textbf{a,b} compares the spectral functions along the $\Gamma$--$H$ path for our atomistic models of defective iron with uncorrelated and $\sim1/\tilde{r}$-correlated vacancies (we recall that we represent the spectral function in the BZ of pristine iron, $\mathcal{B}^{\rm pc}$). 
We see that the distribution of vacancies can dramatically change their effect on the magnon spectral function:
uncorrelated vacancies lead to a very strong blur of the dispersion for $E\gtrsim 350$ meV, while for $\sim1/\tilde{r}$-correlated vacancies the spectral function remains peaked and well-resolved over the entire BZ path. In fact, by fitting a Lorentzian linewidth to the spectral function in this region, the magnon linewidth (Tab.~\ref{table:linewidth}) was found to be between two and three times as large for the spatially uncorrelated vacancies as for the $\sim1/\tilde{r}$-correlated vacancies. 
\begin{table}[!ht]
    \centering
    \begin{tabular}{lcc}
    \toprule
& \multicolumn{2}{c}{Linewidth (meV)} \\
    \cmidrule(lr){2-3}
         $\bm{k}$ ($2\pi/a$) & Uncorrelated & $\sim1/\tilde{r}$ correlated \\
    \midrule
         $(0, 0.51, 0)$ & $107 \pm 14$ & $37 \pm 3$ \\
         $(0, 0.76, 0)$ & $116 \pm 13$ & $57 \pm 6$ \\
    \bottomrule
    \end{tabular}
    \caption{Magnon linewidths of the 1331-atom systems with uncorrelated and $\sim1/\tilde{r}$-correlated vacancies, obtained by fitting a Lorentzian to the spectral function at two wavevectors along the $\Gamma$--$H$ path.}
    \label{table:linewidth}
\end{table}

To connect disorder-induced changes in the magnon spectral function to macroscopic observables, we evaluate the spin stiffness $D$, which is a proxy for the Curie temperature \cite{pajda}, determined as the curvature of the magnon band in the long-wavelength limit,
$\lim_{|\bm{k}|\to 0}E(\bm{k}) {\sim} D|\bm{k}|^2$.
We use the magnon density of states (MDoS) to investigate the direction-averaged spin stiffness (more details in the Methods), which is reported in Table~\ref{table:stiffness}, and whose trend qualitatively agrees with the disorder-driven reduction of spin stiffness discussed in previous experimental \cite{loong,shirane_1968, antonini} and theoretical studies \cite{mudrick,murray}.
The most strongly clustered sample has the lowest fitted stiffness in Table~\ref{table:stiffness}.
To the best of our knowledge, previous experimental studies on the spin stiffness have not quantified the degree of spatial correlation in the distribution of defects; 
the variation across the simulated structures motivates investigations of the spin stiffness and Curie temperature which simultaneously characterize the distribution of vacancies.

\begin{table}[!ht]
    \centering
    \begin{tabular}{lc}
    \toprule
         & $D$ (meV \AA$^2$) \\ 
    \midrule
         Pristine & $331\pm1$  \\
         1331-atom, uncorrelated & $323 \pm 5$  \\
         1331-atom, $\sim26/\tilde{r}$ correlated & $319 \pm 3$  \\
         1000-atom, $\sim76/\tilde{r}$ correlated & $309 \pm 3$  \\
    \bottomrule
    \end{tabular}
    \caption{Direction-averaged spin stiffness $D$ computed from the cumulative MDoS in the low-energy limit. Increasing the prefactor of the $1/\tilde{r}$ correlation corresponds to more clustering and lower fitted stiffness.}
    \label{table:stiffness}
\end{table}

\section*{DISCUSSION}

We introduced a framework leveraging spin-informed machine-learned interatomic potentials to explicitly simulate defective and disordered magnets, featuring distortions in their structure, magnetic moment orientations and magnitudes, and related inhomogeneities in the exchange parameters. We introduced the Coupled Optimization of SpIN Environment (COSINE) algorithm to simultaneously optimize the structural and magnetic degrees of freedom of these systems. In particular, we showed that it is critical to employ this coupled optimizer to converge to a stable configuration, and demonstrated optimizations that make it more robust (with an approximately $30\%$ larger energy drop than conventional MC for the vacancy cluster) and significantly faster (up to seven times) than a per-atom approach. As such, this algorithm could be used to improve the efficiency and robustness of simulation codes such as LAMMPS and other spin-dynamics codes \cite{tranchida,Ma_2016,Skubic_2008,Evans_2014,PhysRevB.99.224414,Bisotti_2018,Dahlbom_2025,Lepadatu_2023}.

We developed a workflow that leverages the magnetic force theorem to compute exchange interactions in materials ranging from insulators to itinerant magnets. We presented an optimization of this workflow for SI-MLIPs with finite receptive field that yields a linear scaling of computational cost with number of atoms, and therefore enables the quantitative simulation of exchange interactions in solids with correlated atomic and nanoscale disorder.
These include compositional inhomogeneities with possible nanoscale correlations, as well as related structural and magnetic distortions, which have hitherto been prohibitively challenging for DFT simulations, due to computational cost scaling cubically with the number of atoms \cite{marzari_electronic_structure}. 
Previous studies dealt with this challenge employing approximations that considered the exchange parameter to be dependent only on chemistry, i.e., neglecting the effect that anisotropies and structural/magnetic distortions can have on it.

We introduced a paraunitary unfolding procedure which generalizes band unfolding to excitations with general bosonic Hamiltonians and paraunitary eigenvectors, which allows us to unfold the magnon spectral function of materials with arbitrary magnetic moment orientations and thus can be applied to antiferromagnets, altermagnets, canted magnets and materials with disorder-induced non-collinearities, and does not require a reference basis to project onto. We use this to obtain the magnon spectral function in full generality, without any assumption on the distribution of the structural defects. 

Compared to established methods, this (i) allows us to fully account for inhomogeneities from the atomic to the nanoscale, and their correlations; (ii) gives us access to mode-resolved quantities which are not directly available and require fitting in spin dynamics \cite{mudrick}. We showed that our approach isolates the disorder-driven contribution to band broadening from other intrinsic sources such as magnon--magnon scattering \cite{zhitomirsky_magnon_decays}. 

Using a spin-informed machine-learned potential for iron \cite{rinaldi}, we demonstrated the workflow by computing the spin stiffness in pristine and defective samples with different correlations in atomic and nanoscale disorder.
We found that, in general, the presence of vacancies attenuates the spin stiffness, with the lowest fitted stiffness observed in the most strongly clustered sample.

This shows the importance of considering not only the concentration, but also the explicit spatial profile of disorder in itinerant magnetic systems, which could be used in the future to tune the material properties and to model third-generation steels \cite{nanda_3rd_gen_ahss, raabe_ahss} or iron subject to neutron irradiation in nuclear applications \cite{beeler}.

Our approach addresses the challenges of computing the spin stiffness of disordered itinerant magnets; the long-range exchange interactions and self-averaging require large supercells \cite{pajda, aharony}, whose explicit simulation is intractable with DFT but becomes accessible with spin-informed MLIPs \cite{chapman_mlip}.

Overall, we introduce the theoretical and computational framework to
systematically explore the capability of emerging spin-informed machine-learning potentials and quantitatively simulate magnons in defective or disordered solids.

We conclude by noting that the theoretical and computational infrastructure we developed is not limited to magnons, and it is generally applicable to any excitation requiring a full bosonic Hamiltonian, including polaritons \cite{kim_polaritons}, triplons \cite{henriques_nanographene}, and Bose--Einstein condensates \cite{jalali}. 

\vspace*{3mm}
\section*{METHODS}

\medskip
\noindent\textbf{Spin Wave Theory}
\nopagebreak

Here, we provide greater detail on the derivation of the second line of Eq.~(\ref{eq:heisenberg_hamiltonian}), using the rotated-frame approach for LSWT first introduced in Ref.~\cite{toth}, and its implementation in \textit{Magnon} \cite{magnon-package}. 

The Holstein--Primakoff transformation ensures that the creation and annihilation operators preserve the spin angular momentum commutation relations \cite{holstein}, and it enforces the physical bounds $|\hat{S}_{\bm{r}b}^{\alpha}|\le S_{\bm{r}b}$. Both follow from the bosonic commutator $[\hat{b}_{\bm{r}b}, \hat{b}_{\bm{r'}b'}^{\dag}] = \delta_{\bm{r}\bm{r'}}\delta_{bb'}$, which defines the number operator $\hat{n}_{\bm{r}b}=\hat{b}_{\bm{r}b}^{\dag}\hat{b}_{\bm{r}b}$. Since a magnon eigenstate is delocalized over the crystal, the expectation value of the number operator of localized spin flips on any given site is small in the low-temperature regime, and the operators are approximated as $\hat{S}^+_{\bm{r}b}\approx\sqrt{2S_{\bm{r}b}}\hat{b}_{\bm{r}b}$ and $\hat{S}^-_{\bm{r}b}\approx\sqrt{2S_{\bm{r}b}}\hat{b}_{\bm{r}b}^{\dag}$ \cite{holstein}, which provides a direct correspondence between lowering the spin and creating a magnon. However, this means that the two aforementioned properties (commutator and bounds) hold only approximately and the eigenstates also become increasingly approximate as the number of excitations is increased. The creation and annihilation operators enter the formulation at first order in $\hat{S}^{x}_{\bm{r}b} = \sqrt{S_{\bm{r}b}/2}(\hat{b}_{\bm{r}b}+\hat{b}^\dag_{\bm{r}b})$ and $\hat{S}^{y}_{\bm{r}b} = i\sqrt{S_{\bm{r}b}/2}(\hat{b}_{\bm{r}b}^\dag-\hat{b}_{\bm{r}b})$, and at second order in $\hat{S}^z_{\bm{r}b}=S_{\bm{r}b}-\hat{n}_{\bm{r}b}$.

Magnons require a consistent quantization axis across the entire material so that the magnetic excitations of any spin configuration may be represented using a single Holstein--Primakoff transformation for every spin, providing a general framework in which no decomposition into sublattices is required \cite{andersonApproximateQuantumTheory1952, kuboSpinWaveTheoryAntiferromagnetics1952}. A mapping is established between the actual spin configuration and a ferromagnetic reference state $\bm{S}_{\bm{r}b}^{\text{FM}}$ (without loss of generality, we take the positive z-direction for the quantization axis). Two rotations provide the mapping, $\bm{S}_{\bm{r}b} = Q_{\bm{r}}Q_{b}\bm{S}^{\text{FM}}_{\bm{r}b}$. In a crystal, represented with a small unit cell, the per-atom rotation $Q_b$ is combined with an additional rotation $Q_{\bm{r}}$ of all the spins in the cell $\bm{r}$ through the same angle, which can describe spin helices and any magnetic ordering commensurate with $\bm{k}\in\mathcal{B}$ \cite{toth}. The rotation $Q_{\bm{r}}$ is absorbed into an effective interaction tensor,
\begin{equation}
    \tilde{J}_{\bm{r}b\bm{r'}b'}^{\alpha\alpha'}= (Q_{\bm{r}}^\top)^{\alpha\beta} J_{\bm{r}b\bm{r'}b'}^{\beta\beta'}Q_{\bm{r'}}^{\beta'\alpha'},
\end{equation}
which must additionally be translation-invariant, $\tilde{J}_{\bm{r}b\bm{r'}b'}^{\alpha\alpha'} = \tilde{J}_{bb'\bm{r''}}^{\alpha\alpha'}$ for $\bm{r'}=\bm{r}+\bm{r''}$.

A magnetic system with disorder, on the other hand, does not necessarily have perfect long-range magnetic order \cite{PhysRevB.38.11718,chapman_mlip,dilucente} nor the same notion of $\bm{k}$-commensurability. $Q_{\bm{r}}$ reduces to the identity, and $Q_b$ describes the long-range magnetic ordering inherited from the parent crystal combined with any disorder-induced deviations. The rotation $Q_{b}$ is broken down into its column vectors, $Q_{b} = (\bm{q}_{b,1}, \bm{q}_{b,2}, \bm{q}_{b,3})$, which are used to express the complex vector $\bm{u}_b = \bm{q}_{b,1} + i\bm{q}_{b,2}$ and real vector $\bm{v}_b = \bm{q}_{b,3}$. Since a consistent convention for the complex phase of the eigenvectors is required for paraunitary unfolding, these vectors must be generated carefully; we use the rotation $Q_b$ to rotate the vectors of the ferromagnetic reference state, $\bm{u}^{\text{FM}}=(1,i,0)$ and $\bm{v}^{\text{FM}}=(0,0,1)$.

In the present work we use periodic boundary conditions for a disordered system represented by a large cell, where their effect is that of a surface term whose contribution decreases relative to the volume as the system size increases, becoming negligible in the thermodynamic limit \cite{ziman}. The cell vector $\bm{r}$ is retained only to represent the exchange interactions of atoms close to the boundary of the supercell. In practice this means it is restricted to the $27$ vectors describing the supercell and its adjacent periodic images, while the overwhelming majority of the interactions between atoms in the interior have $\bm{r}=\bm{0}$. 

When the LSWT Hamiltonian is initially constructed through second quantization and a Fourier transformation, it does not automatically take the form of the second line of Eq.~(\ref{eq:heisenberg_hamiltonian}), but it can instead have any terms bilinear in the operators \cite{colpa}, and there is an apparent freedom to transfer terms between block matrices using the commutator of the creation and annihilation operators. Only the requirement that the Hamiltonian be Hermitian restricts the allowed bilinear operator pairs to the four in Eq.~(\ref{eq:heisenberg_hamiltonian}) and removes the freedom to redistribute them to other pairs by using the commutator. The Hermitian arrangement is found by using the property of a symmetric sum, $\sum_{\bm{k}} f(\bm{k}) = \frac{1}{2}\sum_{\bm{k}}\left[f(\bm{k})+f(-\bm{k})\right]$ to determine the blocks

\begin{widetext}
\begin{equation}
    B_{1bb'}(\bm{k}) = -\frac{1}{2}\sum_{\alpha,\alpha'}\sum_{\bm{r''}}\Bigg\{\frac{\sqrt{S_b S_{b'}}}{2} e^{i\bm{k}\cdot\bm{r''}}e^{i\bm{k}\cdot(\bm{\tau}_{b'}-\bm{\tau}_b)}u^\alpha_b\tilde{J}^{\alpha\alpha'}_{bb'\bm{r''}}\overline{u}^{\alpha'}_{b'}
    +\sum_{b''}-S_{b''}\delta_{bb'}v^\alpha_{b''}\tilde{J}^{\alpha\alpha'}_{b''b'\bm{r''}}v^{\alpha'}_{b'}\Bigg\}
\end{equation}
\begin{equation}
    B_{2bb'}(\bm{k}) = -\frac{1}{2}\sum_{\alpha,\alpha'}\sum_{\bm{r''}}\Bigg\{\frac{\sqrt{S_b S_{b'}}}{2} e^{i\bm{k}\cdot\bm{r''}}e^{i\bm{k}\cdot(\bm{\tau}_{b'}-\bm{\tau}_b)}u^\alpha_b\tilde{J}^{\alpha\alpha'}_{bb'\bm{r''}}u^{\alpha'}_{b'}\Bigg\}
\end{equation}
\end{widetext}

where the complex vector $u^\alpha_b$ determines which block the interactions are placed upon in the Hamiltonian based on the relative orientation of the two interacting magnetic moments. 

\medskip
\noindent\textbf{Paraunitary Diagonalization}
\nopagebreak

The diagonalizing transformation of the Hamiltonian must ensure that the diagonalizing basis of creation and annihilation operators preserves the commutators of the original creation and annihilation operators, which requires that the eigenvectors satisfy Eq.~(\ref{eq:paraunitary_condition}), the defining property of paraunitary eigenvectors. To enforce this condition when performing the diagonalization, the algorithm according to Ref.~\cite{colpa} is used:

\begin{enumerate}
    \item Writing $B(\bm{k})$ for the block matrix in Eq.~(\ref{eq:heisenberg_hamiltonian}), we Cholesky-decompose $2B(\bm{k})=\mathcal{C}^\dag \mathcal{C}$. The factor of two accounts for the full Nambu sum in that equation.
    \item We find the unitary matrix $\mathcal{U}$ which diagonalizes $\mathcal{C} \Phi \mathcal{C}^\dag$, i.e., $\mathcal{U}^\dag[\mathcal{C}\Phi\mathcal{C}^\dag]\mathcal{U} = \mathcal{L}$, a diagonal matrix.
    \item The diagonal matrix of eigenenergies of the original Hamiltonian $\mathcal{E} = \Phi\mathcal{L}$, and the paraunitary eigenvectors $\mathcal{T} = \mathcal{E}^{-\frac{1}{2}}\mathcal{U}^{-1}\mathcal{C}$
\end{enumerate}

The Cholesky decomposition requires that the Hamiltonian is positive-definite. 

\medskip
\noindent\textbf{Symmetries of the exchange interactions}
\nopagebreak

Two main symmetries should be satisfied by the exchange interactions; first, from a physical standpoint, they should be consistent with the underlying symmetry of the crystal, and second, each interaction should be invariant under exchange of the two sites,

\begin{equation}
    \label{eq:bond_inversion_symmetry}
    J^{\alpha'\alpha}_{b'b\bm{r''}}=J^{\alpha\alpha'}_{bb'-\bm{r''}},
\end{equation}

which we term `bond-inversion symmetry'. The latter is a prerequisite for the Hamiltonian in Eq.~(\ref{eq:heisenberg_hamiltonian}) to be Hermitian \cite{toth}. The \textit{Magnon} software uses Spglib \cite{spglib} to obtain magnetic point-group symmetry operations and can generate full sets of exchange interactions for structures using both types of symmetry, starting from a minimum descriptive set from the literature or identifying inconsistencies in calculated exchange interactions. It is critical that nonmagnetic sites are included in the unit cell, because even though they may not contribute directly to the magnon spectrum, their absence can result in the incorrect symmetry group being identified. In some systems, truncating the exchange interactions at too small a shell can spuriously increase the symmetry of the exchange interactions, so it is important to include sufficiently many interactions to avoid this \cite{gohlkeSpuriousSymmetryEnhancement2023a}.

\medskip
\noindent\textbf{Band Unfolding}
\nopagebreak

In this section, we discuss how the eigenvectors of a pristine system with translational symmetry exhibit definite signatures of this symmetry, and explain how this observation can be used to perform band unfolding without a reference basis to project onto, using the insights of Ref.~\cite{allen_unfolding}. 

A cell with periodic boundary conditions represents a bulk solid, and there are many possible choices of cell which, when repeated, will yield the same structure. A fundamental effect of repeating the primitive unit cell to create a `supercell' is the exchange of information between real and reciprocal space, or equivalently between eigenvectors and the Bloch phase. Significantly, the physical amplitude $\Psi_{\bm{k}\bm{R}\bm{r}\beta\eta}$ of a certain magnon excitation on each atom should be identical in all possible representations via different periodic cells; because the supercell is pristine, we replaced the atom index $b$ with $\bm{r}$ labeling copies of the primitive cell within the supercell and the atom index $\beta$ in the primitive cell. Then, in terms of the supercell eigenvector,
\begin{equation} \label{bloch_supercell}
    \Psi_{\bm{k}\bm{R}\bm{r}\beta\eta} = \mathcal{T}^{\rm sc}_{\bm{k}\bm{r}\beta\eta} e^{i\bm{K}\cdot(\bm{R}+\bm{r}+\bm{\tau}_\beta)},
\end{equation}
since the physical amplitude is a Bloch function of the supercell. The role of $\bm{K}$ is to provide a plane wave modulation of the supercell eigenvector, enabling Fourier sampling of the entire bulk sample.

Without loss of generality, we consider the example of a perfect $\mathcal{N}_{\rm c}\times1\times1$ supercell, in which case there is an additional symmetry under translations by lattice vectors of the original $1\times1\times1$ cell $r'^{x} = R^{x}/\mathcal{N}_{\rm c}$. Specifically, $\Psi_{\bm{k}\bm{R}\bm{r}\beta\eta}$ must also be a Bloch function of the primitive lattice, with the property that
\begin{equation} \label{bloch_primitive}
    \hat{T}_{\bm{r'}} \Psi_{\bm{k}\bm{R}\bm{r}\beta\eta} = e^{i(\bm{G}+\bm{K})\cdot\bm{r'}}\Psi_{\bm{k}\bm{R}\bm{r}\beta\eta},
\end{equation}
where $\hat{T}_{\bm{r'}}\Psi_{\bm{k}\bm{R}\bm{r}\beta\eta} = \Psi_{\bm{k}\bm{R}(\bm{r}+\bm{r'})\beta\eta}$. By combining Eq.~(\ref{bloch_supercell}) and Eq.~(\ref{bloch_primitive}), it is straightforward to show that the supercell eigenvectors translated by a primitive lattice vector differ by a phase due to supercell reciprocal lattice vectors $\bm{G}$ only,
\begin{equation} \label{eq:eigenvector_translation}
    \mathcal{T}^{\rm sc}_{\bm{k}(\bm{r}+\bm{r'})\beta\eta} = e^{i\bm{G}\cdot\bm{r'}} \mathcal{T}^{\rm sc}_{\bm{k}(\bm{r})\beta\eta}.
\end{equation}
In other words, in order that Bloch's theorem holds identically in both representations of the same pristine crystal, the supercell eigenvectors must have the property of Eq.~(\ref{eq:eigenvector_translation}) under translations. The requirement that
\begin{equation}
    (\hat{T}_{\bm{r}})^{\mathcal{N}_{\rm c}} = \hat{I},
\end{equation}
the identity operator, also means that the eigenvector elements which map between each other under translations differ only by factors which are the $\mathcal{N}_{\rm c}$th roots of unity, $e^{\frac{i2\pi n}{\mathcal{N}_{\rm c}}}$, for $n=0,\ldots,\mathcal{N}_{\rm c}-1$. In other words, the translation operators form an Abelian group for which the representation is the difference in phase between translation-equivalent eigenvector elements \cite{togo_unfolding}. The act of creating the supercell reduces the volume of the BZ by a factor of $\mathcal{N}_{\rm c}$, restricting the $\bm{k}=\bm{G}+\bm{K}$ to the values where $\bm{G}=\bm{0}$ i.e., a smaller volume in reciprocal space which can only describe plane wave modulations of the supercell, not of the original primitive cell (i.e., the Nyquist frequency is smaller for a bigger supercell). However, the $\mathcal{N}_{\rm c}$ roots of unity provide $\mathcal{N}_{\rm c}$ choices for the phase difference between the eigenvector components, which provide an additional modulation of $e^{i\bm{G}\cdot \bm{r}}$. The additivity of wavevectors,
$ e^{i\bm{G}\cdot \bm{r}}e^{i\bm{K}\cdot \bm{r}} = e^{i(\bm{G}+\bm{K})\cdot \bm{r}}$
provides access to all of the possible modulations available in the original BZ of the primitive cell, ensuring that they are identical representations. In effect, by constructing the supercell, some of the complex phase has been `transferred' to the eigenvectors. The bandstructure of the supercell at $\bm{K}$ then contains $\mathcal{N}_{\rm c}$ times as many bands because there are $\mathcal{N}_{\rm c}$ unique modulations $e^{i\bm{G}\cdot \bm{r}}$ which can be applied to reach $\mathcal{N}_{\rm c}$ different $\bm{k}$ from each $\bm{K}$. Therefore, when unfolding we perform an inverse Fourier transform over the components of the eigenvectors using the permutations in Eq.~(\ref{eq:permutations}). In a disordered system, provided that a one-to-one mapping can be established between real-space positions of the atoms of the disordered system and those of a pristine supercell (such as when the displacements of the atomic positions are small \cite{togo_unfolding}) the approximate translational symmetry of the eigenvectors means that this method can still be used, but the spectral function will be blurred. 

To represent the eigenvalues of the magnon Hamiltonian of the defective system in the BZ of the parent crystal, we represented a vacancy (both in the structural and magnetic sense) as a `ghost' atom and spin which have no interactions with any other atom and spin, adding a zero row and column in each Nambu sector to Eq.~(\ref{eq:heisenberg_hamiltonian}) and introducing a vacancy zero mode that must be excluded from the physical MDoS while preserving the one-to-one mapping. Our approach can be considered as a magnetic extension of the approach developed to describe structural vacancies in Ref.~\cite{togo_unfolding}.

\medskip
\noindent\textbf{Generation of disordered supercells}
\nopagebreak

\begin{figure}[!ht]
    \centering
    \includegraphics[width=\columnwidth]{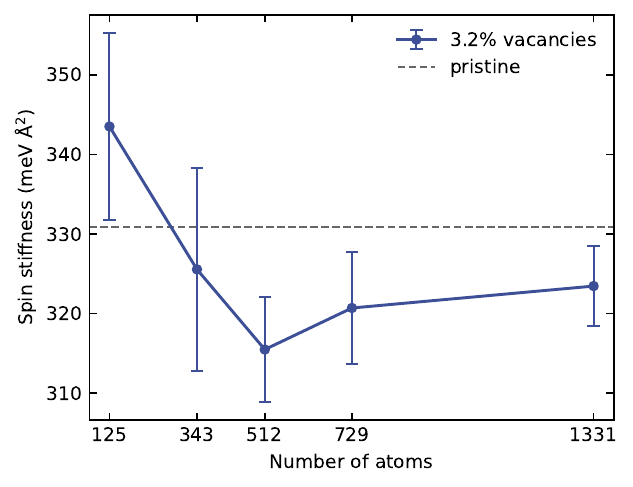}
    \caption{The spin stiffness obtained from the cumulative MDoS of systems generated with spatially uncorrelated vacancies and varying numbers of atoms.}
    \label{fig:stiffness_convergence}
\end{figure}

The defective supercells with non-trivial spatially correlated defects were constructed using a correlated noise field which was generated by making use of the convolution theorem to combine the correlation function $C(\tilde{r})$ in Eq.~(\ref{eq:correlator}) with random noise \cite{thebaud, fiorentino}. Point defects were added iteratively to lattice sites at the extrema of the correlated noise field. The realized spatial correlations of vacancies were then quantified by calculating $C(\tilde{r})$ and fitting a general inverse power law to extract the coefficient, which determines the strength of clustering, and the rate of decay of spatial correlations. This requires performing convergence tests of the target properties against different supercell sizes; as shown in Fig.~\ref{fig:stiffness_convergence}, the direction-averaged spin stiffness obtained from systems with spatially uncorrelated vacancies converges, within uncertainty, for systems with $N_{\rm at}\gtrsim343$. 

\medskip
\noindent\textbf{Coupled Optimization of SpIN Environment}
\nopagebreak

The COSINE algorithm synchronizes MC updates across magnetic moments with symmetry-equivalent LAEs. This restricts the search space and can exclude symmetry-breaking relaxations. The multiple initializations and comparison with per-atom MC described above test this restriction for the structures studied.

We employ several techniques to improve efficiency. Energy evaluations are performed with only the relevant subsystem whereby atoms at a distance of more than $2r_c$ from any of the atoms in the perturbed group are removed, and the energy difference is computed only on the remaining atoms. Any atoms within $r_c$ are directly affected by the perturbation, and any atoms further than this but closer than $2r_c$ are affected indirectly as they share cluster atoms within the $r_c$ region. The total energy updated in this way is periodically refreshed and the energy `slip' is verified to be within numerical precision. 

During structural relaxation calculations, we used an adaptive annealing procedure whereby the step size was periodically updated based on the running acceptance ratio for each type of move (e.g., magnetic moment magnitudes, polar and azimuthal angles, and cell vector lengths) \cite{Alzate_Cardona_2019}, followed by a final validation period with a larger step size to check convergence. This ensures that, given the roughness of the mPES, the structure is not arbitrarily forced into a local minimum by a rigid annealing schedule.

\medskip
\noindent\textbf{Tilted Spins Method}
\nopagebreak

\begin{figure}[!ht]
    \centering
    \includegraphics[width=\columnwidth]{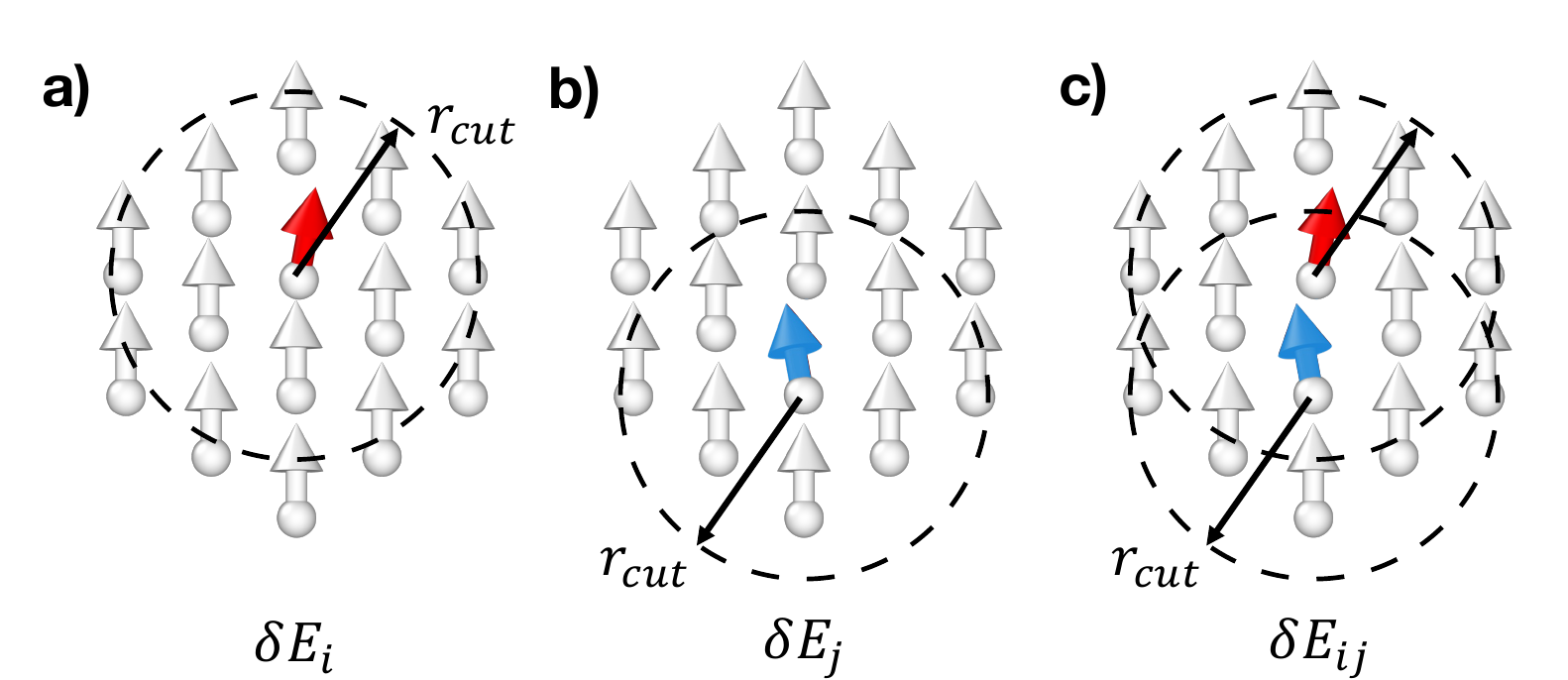}
    \caption{The energies required to predict the exchange interaction between two magnetic sites $i$ and $j$, in red and blue, respectively. The calculation of the energy difference requires that each evaluation considers all atoms within $r_c$ of either atom. (a)--(c) represent the three perturbations related to the energies $\delta E_i$, $\delta E_j$ and $\delta E_{ij}$.}
    \label{fig:liechtenstein}
\end{figure}

The Heisenberg Hamiltonian Eq.~(\ref{eq:heisenberg_hamiltonian}) is a phenomenological Hamiltonian which labels the magnetic contribution to the total energy in terms of the relative orientations of the spin magnetic moments \cite{heisenberg} and the coefficient of each pairwise interaction may be extracted from a series of small tilts of the magnetic moments to probe this contribution \cite{liechtenstein}. We calculate the energy changes $\delta E_i$, $\delta E_j$ and $\delta E_{ij}$ of the tilts shown in Fig.~\ref{fig:liechtenstein}, where $i$ and $j$ tilt through an angle $\theta$ in opposite directions, which can be done directly due to the computational speed of SI-MLIPs \cite{rinaldi}, whereas DFT studies based on the magnetic force theorem usually use a perturbation theory approach on top of collinear calculations because of the significant expense of non-collinear calculations. Alternative methods include total energy mapping approaches, time-dependent DFT, and DFT calculations of structures with magnons condensed into the spin configuration \cite{Rezaei_2026, RevModPhys.95.035004}. Energy mapping approaches, which use different collinear spin configurations, are unsuitable in itinerant magnetic materials where the exchange interactions can be configuration dependent \cite{kvashninMicroscopicOriginHeisenberg2016,szilvaTheoryNoncollinearInteractions2017}. The computational expense of time-dependent and constrained non-collinear DFT is prohibitive for large disordered cells \cite{marzari_electronic_structure, horton_magnetic_order, li_deeph_magnetic}.

The \textit{Magnon} code \cite{magnon-package} introduces massively parallel exchange interaction calculators to compute large datasets of interactions efficiently and is agnostic to the energy calculator, allowing it to be interfaced with DFT and SI-MLIP calculations. SI-MLIPs are trained on DFT data to produce a model with comparable quantum accuracy but faster evaluation \cite{pan_scaling, kalita}, and can incorporate equivariance so that physical symmetries are preserved in the internal structure of the model \cite{kalita}. The non-collinear SI-MLIP for iron used in this work \cite{rinaldi} has a radial cutoff $r_c=4.5$~\AA; since the primitive unit cell of a BCC crystal is non-orthogonal \cite{Stevenson_2021} (which means that its inradius is smaller than half of the smallest cell length), the training data will only enable accurate fitting up to a distance set by the inradius of its cell, even though two atoms can interact up to a separation distance of $2r_c$. Within this constraint, the SI-MLIP allows predictions of exchange interactions up to the sixth-nearest neighbor.

\medskip
\noindent\textbf{Unit spin model}
\nopagebreak

There exist two different conventions in the literature for the magnitudes of the spins. The Hamiltonian $H_S$ in Eq.~(\ref{eq:heisenberg_hamiltonian}) uses spin operators in units of $\hbar$; alternatively, a unit spin model $\hat{H}_s$ with unit vectors $\hat{\bm{s}}_{\bm{r}b}$ is often used \cite{Evans_2014, PhysRevB.99.224414, Mentink_2010, Radu_2011}, in which case the magnitudes are absorbed into the definition of the exchange interaction. With $\hat{\bm{S}}_b=S_b\hat{\bm{s}}_b$, the intersite quadratic Hamiltonian matrix elements obtained using spin lengths $S_b$ and unity, respectively, are related by \cite{chapman_nonheisenberg}

\begin{equation}
(\hat{H}_S^{(2)})_{bb'} \propto \sqrt{\frac{1}{S_{b}S_{b'}}} (\hat{H}_s^{(2)})_{bb'},
\label{eq:unit_spin_model}
\end{equation}

where the exact factor depends on differences in conventions between the two Hamiltonians such as the prefactor (i.e., whether or not there is double counting in the sum) and sign convention for ferromagnetic exchange. The transformation is implemented in \textit{Magnon} \cite{magnon-package}.

\medskip
\noindent\textbf{Spin Stiffness}
\nopagebreak

In metallic magnetic systems, there has been large disagreement between the spin stiffness obtained using varying methods including bandstructure fitting, real-space summation and spin-spiral methods \cite{siprAssessingDifferentApproaches2020}. Contributions to the variance include the long-range nature of the exchange interactions, which make real-space summation difficult to converge and require dense meshes in DFT calculations \cite{pajda}. Stiffness values obtained from fitting to the magnon dispersion are sensitive to higher-order terms in $\bm{k}$, and methods using spin spirals or the magnetic force theorem should ensure that the obtained values do not depend on the tilting angle and choice of $\bm{k}$ used in the fit.

The magnitudes of the exchange interaction coefficients are known to decay with distance \cite{sipr, kvashnin_dmft_exchange, katsnelson_correlated_exchange}, but unlike insulators, in which this decay is exponential, itinerant magnets can display $r^{-3}$ scaling \cite{sipr, kvashninMicroscopicOriginHeisenberg2016} due to RKKY, which requires convergence of the spin stiffness against the number of shells \cite{pajda}. The size of the supercell required to predict a beyond-mean-field exchange interaction scales with the pairwise atomic separation, which would make the calculations of full exchange datasets in disordered materials intractable with DFT \cite{marzari_electronic_structure, horton_magnetic_order, li_deeph_magnetic}. 

For one three-dimensional quadratic magnon branch, the MDoS per unit volume $G(E)=\sqrt{E}/(4\pi^2D^{\frac{3}{2}})$, and from the cumulative MDoS $\mathcal{N}(E) = \int_{0}^{E} G(E')\,\mathrm{d}E'$ (the number of modes with energy lower than $E$ per unit volume) the stiffness
\begin{equation}
    D = \frac{E}{(6\pi^2\mathcal{N}(E))^{2/3}}
\end{equation}
is calculated. The direction-averaged spin stiffness computed in this way equals $(\det\mathbf{D})^{1/3}$ for a quadratic dispersion $E=\bm{k}^{\mathsf{T}}\mathbf{D}\bm{k}$. Directional variations were found to be $\lesssim10$ meV \AA$^2$ for the correlated samples. The range of $E$ over which to perform the fit was determined by converging the error in the fitted values against an increasing upper limit, since the uncertainty is at first large due to the small number of points.

\noindent
\textbf{Data availability.}
The magnetic ACE potential for iron is available at https://github.com/mttrin93/mag-ace/tree/master, and its training data is online at https://zenodo.org/records/15363836.
\noindent
\textbf{Code availability.}
The code used to perform the LSWT calculations, COSINE for magnetic structures, paraunitary unfolding of magnon bands, spectral function calculation and exchange interaction calculations will be made available in the next release of the \textit{Magnon} Python library \cite{magnon-package}, available at https://github.com/MPA2suite/Magnon and on the Python Package Index at https://pypi.org/project/magnon/.

\vspace*{2mm}
\noindent
\textbf{Acknowledgements.}
This material is based upon work supported by the NSF MRSEC program through Columbia in the Center for Advanced Electronic Materials (DMR-2614607). It also used the ARCHER2 UK National Supercomputing Service (https://www.archer2.ac.uk) and computing resources from Columbia University's Shared Research Computing Facility. A.M. acknowledges support from the Harding Distinguished Postgraduate Scholarship in the early phase of this work.

\noindent
\textbf{Author contributions.}
M.S. conceived and directed the project. 
A.M. performed the analytical derivations, wrote the magnon simulation code, and performed the numerical simulations, with inputs from J.B. and M.S. 
A.M. prepared the figures and wrote the first draft of the manuscript, and revised it with contributions from M.S. All authors reviewed and commented on the manuscript.

\noindent
\textbf{Competing interests.}
All authors declare no financial or non-financial competing interests. 

\noindent
\textbf{Inclusion \& ethics statement.}
All the authors of this study have fulfilled the criteria for authorship required by Nature Portfolio journals, and their participation was essential for the design and implementation of the study. Roles and responsibilities were discussed and agreed among collaborators. Local and regional research relevant to our study has been considered in the citations.

%

\end{document}